\documentclass[prd,showpacs,showkeys,preprintnumbers,floatfix,
nofootinbib,superscriptaddress]{revtex4-1}

\usepackage{slashed}
\usepackage{mathtools}
\usepackage{amsfonts} 
\usepackage{amssymb} 
\usepackage{amsmath} 
\usepackage{graphicx} 
\usepackage{subfigure} 
\usepackage{array} 
\usepackage{dcolumn} 
\usepackage{bm} 
\usepackage{latexsym} 
\usepackage{longtable} 
\usepackage{hyperref} 
\usepackage{verbatim}
\usepackage{epsfig}
\usepackage{color}

\def\tilde{\widetilde}

\newcommand{\tMi}[1]{\widetilde{\mathcal M}_{2,#1}}

\newcommand{\M}[0]{\mathcal M_2}
\newcommand{\K}[0]{\mathcal K_2}

\newcommand{\tK}[0]{\widetilde{\mathcal K}_2}
\newcommand{\tM}[0]{\widetilde{\mathcal M}_2}
\newcommand{\Kin}[1]{\mathcal K_{2; #1}}
\newcommand{\tMin}[1]{\widetilde{\mathcal M}_{2; #1}}
\newcommand{\tKin}[1]{\widetilde{\mathcal K}_{2; #1}}
\newcommand{\Kdf}[0]{\mathcal K_{\mathrm{df},3}}

\newcommand{\tKdf}[0]{\widetilde{\mathcal K}_{\mathrm{df},3}}
\newcommand{\Mdf}[0]{\mathcal M_{\mathrm{df},3}}
\newcommand{\Mdfin}[1]{\mathcal M_{\mathrm{df},3;#1}}
\newcommand{\Kdfin}[1]{\mathcal K_{\mathrm{df},3;#1}}
\newcommand{\tKdfin}[1]{\widetilde{\mathcal K}_{\mathrm{df},3;#1}}

\newcommand{\Mth}[0]{\mathcal M_3}
\newcommand{\Mthr}[0]{\mathcal M_{3,\mathrm{th}}}
\newcommand{\DEth}[0]{\Delta E_{\rm th}}
\newcommand{\Rzero}[0]{\mathcal R_0}

\begin{document}

\title{Threshold expansion of the three-particle quantization condition}
\author{Maxwell T. Hansen}
\email[Email: ]{hansen@kph.uni-mainz.de}
\affiliation{Institut f\"ur Kernphysik and Helmholz Institute Mainz, Johannes Gutenberg-Universit\"at Mainz,
55099 Mainz, Germany\\
}
\author{Stephen R. Sharpe}
\email[Email: ]{srsharpe@uw.edu}
\affiliation{
 Physics Department, University of Washington, 
 Seattle, WA 98195-1560, USA \\
}
\date{\today}
\begin{abstract}
We recently derived a quantization condition for the
energy of three relativistic particles in a cubic box~\cite{us,KtoM}.
Here we use this condition to 
study the energy level closest to the three-particle threshold
when the total three-momentum vanishes.
We expand this energy in powers of $1/L$, where $L$ is the
linear extent of the finite volume.  The expansion begins at ${\cal O}(1/L^3)$, 
and we determine the coefficients of the terms through
${\cal O}(1/L^6)$. As is also the case for the two-particle threshold
energy, the $1/L^3$, $1/L^4$ and $1/L^5$ coefficients depend only on
the two-particle scattering length $a$. These can be compared to
previous results obtained using nonrelativistic quantum
mechanics~\cite{Huang,Beane2007,Tan2007}, and we find complete
agreement.  The $1/L^6$ coefficients depend additionally on the
two-particle effective range $r$ (just as in the two-particle case)
and on a suitably defined threshold three-particle scattering amplitude (a new feature for three particles). A second new feature in the three-particle case is that logarithmic dependence on $L$
appears at $\mathcal O(1/L^6)$.  Relativistic effects enter at this
order, and the only comparison possible with the nonrelativistic result
is for the coefficient of the logarithm, where we again find
agreement.  For a more thorough check of the $1/L^6$ result, and thus of the quantization condition, we also compare to a perturbative
calculation of the threshold energy in relativistic $\lambda \phi^4$
theory, which we have recently presented in Ref.~\cite{ourpt}. Here,
all terms can be compared, and we find full agreement.
\end{abstract}
\maketitle


\section{Introduction}
\label{sec:intro}

In two recent papers, we
derived a relation between the spectrum of three
relativistic particles in a periodic box and on-shell, infinite-volume two-to-two and three-to-three scattering
amplitudes~\cite{us,KtoM}. In the first paper, Ref.~\cite{us}, we related the finite-volume spectrum to an unphysical infinite-volume three-to-three scattering quantity that we denoted $\Kdf$. The formalism was then completed in Ref.~\cite{KtoM}, where we presented the purely infinite-volume relation between $\Kdf$ and the standard three-to-three scattering amplitude, $\mathcal M_3$.
As the derivation of these results is lengthy and involved, it is important to
check them as thoroughly as possible.
Some checks were made in Refs.~\cite{us,KtoM}, but
the purpose of the present paper is to provide a more significant check.
We do so by calculating, in our formalism, the energy of the state closest to threshold
as a function of the inverse box size $1/L$,
and by comparing to results obtained using two other methods:
nonrelativistic quantum mechanics (NRQM)
(as done in Refs.~\cite{Huang,Beane2007,Tan2007})
and a perturbative expansion in relativistic $\lambda\phi^4$ theory
(a calculation we have recently completed in Ref.~\cite{ourpt}).
These two methods provide complementary checks of the results of our general formalism.

The result derived in Refs.~\cite{us,KtoM} is for a scalar field $\phi$
with a $Z_2$ symmetry, $\phi \rightarrow - \phi$, so that only
even legged vertices appear. This theory is studied in 
a cubic box with side length $L$ and periodic boundary conditions 
in all three spatial directions. The absence of $2\longrightarrow 3$
transitions means that a direct comparison can be made to the nonrelativistic approach,
since in the latter particle number is conserved.

The analysis of Refs.~\cite{us,KtoM} allows for nonzero total three-momentum, 
$\vec P$, in the finite-volume frame. 
However, since Refs.~\cite{Huang, Beane2007,Tan2007,ourpt}
consider only zero total three-momentum, we restrict ourselves here
to $\vec P=0$.
This means that the threshold occurs when
the total energy satisfies $E=3m$,
with $m$ the physical mass of the scalar particle.
In the absence of interactions, this is also the energy of the lowest-lying
three-particle state in the box, with all particles at rest.
Including interactions, the energy of this state will shift by an amount 
\begin{equation}
\label{eq:DeltaEdef}
\DEth = E -3m\,,
\end{equation}
which should go to zero as $L\to\infty$.
For two particles, it is well known that 
$\DEth \propto a/L^3 + {\cal O}(1/L^4)$,
with $a$ the scattering length 
(see Ref.~\cite{Luscher:1986n2} and references therein).
The $1/L^3$ factor 
arises because the two particles, both of which have
spatially uniform wave functions, need to be close to each other in order to interact.  
We expect that $\DEth$ for three particles should scale with the
same power of $1/L$, since one possible process is a pairwise interaction 
with the third particle spectating. 
Similarly, a localized three-particle
interaction should lead to a contribution scaling as $1/L^6$,
since all three particles must be close.
These expectations are indeed borne out by the results of 
Refs.~\cite{Beane2007,Tan2007,ourpt}.

It should be noted that, in finite volume, there is an infinite tower of states with energies $E_n(L)$
satisfying $\lim_{L \rightarrow \infty} E_n(L) = 3 m$.
We are only interested in the lowest lying level in this infinite set, for which, as noted above,
$\Delta E={\cal O}(1/L^3)$.
In particular we are not concerned with excited states, that, in the noninteracting limit, contain at least two 
particles with nonzero momenta. The energy shifts for such states scale as
$\Delta E={\cal O}(1/L^2)$ with positive coefficients.
Our quantization condition could also be used to develop the $1/L$ expansion of 
the energy shifts for these excited states, but we do not pursue this in the present article.  

In light of these considerations we expand the energy shift as,
\begin{equation}
 \DEth = \sum_{n=3}^\infty \frac{a_n(L)}{L^{n}} \,,
\label{eq:DeltaEexp}
\end{equation}
and determine the $a_n(L)$ up to $n=6$. 
We include a possible $L$ dependence in the coefficients,
since Refs.~\cite{Beane2007,Tan2007,ourpt} find a logarithmic
dependence for $a_6(L)$.

As we will show, our results for $a_{3-5}$, as well as the logarithmic, volume-dependent
term in $a_6(L)$, agree with those from Refs.~\cite{Huang,Beane2007,Tan2007}
(which were also checked in Ref.~\cite{ourpt}).
We cannot, however, make a useful comparison with the NRQM results for the 
volume-independent part of $a_6(L)$. 
This is for two reasons. First, as we discovered in Ref.~\cite{ourpt}, there are differences
between the nonrelativistic and relativistic results for the two-particle threshold energy shift
at ${\cal O}(1/L^6)$. Such differences arise from relativistic kinematics, and we expect these to
persist also in the three-particle case.
Second, this is the order at which a three-particle interaction first appears,
and the definition of this quantity is scheme dependent.
The schemes used in the two NRQM calculations differ from that used
in our formalism (as well as from each other), and the relationship between
these schemes is not known at present.
It is primarily because of this issue that we carried out the perturbative calculation of Ref.~\cite{ourpt},
since in that calculation we could use the same scheme for defining the three-particle interaction,
and thus provide an unambiguous check for $a_6(L)$.\footnote{See also Ref.~\cite{maxpro} for a recent review of results for three particles in a finite volume.}

\begin{figure}
\begin{center}
\includegraphics[scale=0.38]{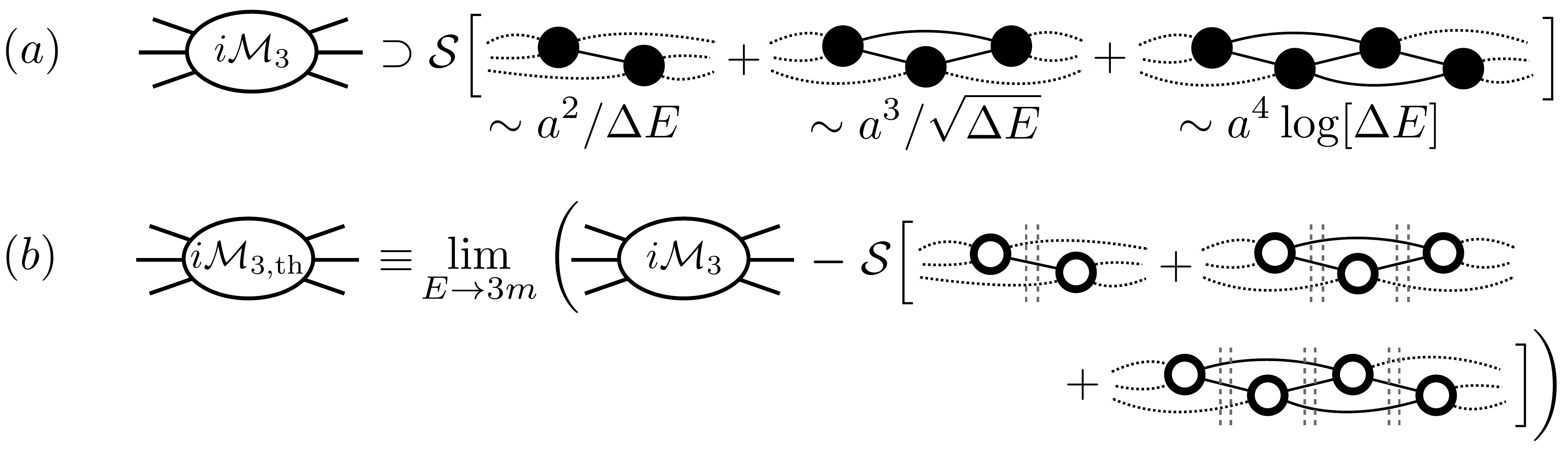}
\end{center}
\caption{(a) The three-to-three scattering amplitude contains three types 
of pairwise scattering diagrams that lead to divergences at threshold. 
The scaling of the divergence with the shift from threshold, $\Delta E$, 
and the two-particle scattering length, $a$, is shown. 
(b) We define a finite threshold scattering amplitude by subtracting the 
singular parts of these diagrams before sending the energy to $3m$. 
The rings, in contrast to filled circles, indicate that only the two-to-two scattering 
amplitude near threshold appears in the subtraction. 
The vertical dashed lines indicate that a simple pole is used in place of the fully dressed propagator.
Detailed definitions are given in Sec.~\ref{sec:MdftoMthr}.}
\label{fig:Mthr}
\end{figure}

Since the scheme dependence of the three-particle interaction plays an important
role in the following, we briefly recall how this issue arises.
The quantity that one naively expects to enter the $1/L^6$ energy shift in a relativistic theory
is the infinite-volume three-to-three scattering amplitude at threshold. 
This cannot be the case, however, since this amplitude diverges as $\Delta E=E-3m$ vanishes.\footnote{%
As discussed later, there are also divergences above threshold. We imagine here choosing the
kinematics such that the above-threshold divergences are avoided, and then moving towards threshold,
at which point the divergences cannot be avoided.}
The divergences are due to the three pairwise scattering diagrams shown in Fig.~\ref{fig:Mthr}(a),
and scale as $a^2/\Delta E$, $a^3/\sqrt{\Delta E}$ and $a^4\log(\Delta E)$, respectively,
where $a$ is the scattering length.
The existence of such singularities is a general field-theoretic result that
was established long ago~\cite{Rubin:1966zz,Brayshaw:1969ab,Taylor:1977A,Taylor:1977B}. 
Our formalism accommodates these divergences by 
finding that $\DEth$ depends on a modified quantity, $\Mthr$, 
given by subtracting the divergent terms from $\Mth$ 
[see Fig.~\ref{fig:Mthr}(b) as well as Eq.~(\ref{eq:Mthrdef}) below]. 
The choice of subtraction is, however, ambiguous, and introduces dependence
on a cutoff scale and scheme.

The remainder of this paper is organized as follows. 
In the following section we summarize the quantization condition of Ref.~\cite{us}, 
which takes the form of a determinant of formally infinite-dimensional matrices. 
The core of this paper is Sec.~\ref{sec:thr}, in which we describe the development 
of the threshold expansion. 
The central difficulty is that, at $\mathcal O(1/L^6)$ in the expansion of $\DEth$,
 all entries in the infinite-dimensional matrices contribute. 
 We thus first recast the quantization condition into a more useful form, 
given in Eq.~(\ref{eq:qcondnomat}). 
 We then analyze the reduced result by understanding the $1/L$ scaling of its components.
 The analysis  is rather involved and lengthy,
 and requires the introduction of the threshold amplitude $\Mthr$ discussed above.
 Brief conclusions are given in Sec.~\ref{sec:conc}. 
 Technical calculations are collected in three appendices.


\section{Summary of quantization condition}
\label{sec:qc}

In this section we recall the three-particle quantization condition 
from Ref.~\cite{us}. This condition determines
the spectral energies, $E_i$, to be those values for which
\begin{equation}
\label{eq:qcond}
\det [1 + F_{3} \Kdf]  =  0 \,.
\end{equation}
Here $F_3$ and $\Kdf$ are matrices, to be defined below,
that depend on $E$ (and, in the case of $F_3$, also on $L$).
Particle interactions enter through two
infinite-volume scattering quantities: 
the three-particle quantity $\Kdf$, shown explicitly, 
and the two-particle K matrix $\K$, contained in $F_3$.

$\Kdf$ is a three-particle divergence-free K matrix. 
It depends on the same (on shell) kinematic variables as ${\cal M}_3$,
and is invariant under interchange of the external particle momenta.
It differs from the standard three-to-three scattering amplitude
in two important ways~\cite{us,KtoM}. First, physical  divergences, 
which are known to occur in the three-particle 
scattering amplitude, ${\cal M}_3$, are absent in $\Kdf$. 
These divergences are due to pairwise scatterings separated 
by arbitrarily long lived  intermediate states [see Fig.~\ref{fig:Mthr}(a)].
Second, loop integrals defining $\Kdf$ are evaluated with 
a pole prescription differing from the standard $i\epsilon$ prescription.
This feature is needed to properly accommodate finite-volume effects from 
the two-particle unitary cusp. The issue of two-particle cusps plays a minor role 
in the threshold expansion so we do not describe it here. 
We direct the interested reader to Refs.~\cite{us,KtoM,maxpro} for a thorough discussion.

The precise relation between $\Kdf$ and $\Mth$ is given in Ref.~\cite{KtoM}. 
First, one uses an integral equation to convert $\Kdf$ to $\Mdf$. 
The latter is an intermediate quantity that, like $\Kdf$, 
has no singularities due to long-lived intermediate states. 
Unlike $\Kdf$, however, $\Mdf$ is defined with the standard $i \epsilon$-pole prescription 
and is therefore more closely related to the standard scattering amplitude. 
$\Mdf$ is defined in Eq.~(\ref{eq:Mdfdef}) below.
Second, one adds back in the singular terms. 
These depend only on kinematic variables as well as 
the on-shell two-to-two scattering amplitude. As we see below, 
the threshold expansion of Eq.~(\ref{eq:qcond}) actually reproduces the integral equation that 
converts $\Kdf$ to $\Mdf$. In addition, the expansion produces an infinite series of terms 
that convert $\Mdf$ to the quantity $\Mthr$ introduced above. Of the three quantities, $\Kdf$, $\Mdf$ and $\Mthr$, only the latter appears in our final result for the threshold expansion. This is also the quantity that is most closely related to the standard scattering amplitude.

We now explain the matrix indices of $\Kdf$ and $F_3$.
These specify the incoming and outgoing configuration of three
on-shell particles with $\vec P=0$ and given total energy $E$.
We arbitrarily pick one of the three incoming particles and label
its momentum $\vec k$, and similarly label one of the outgoing
momenta $\vec k'$. We sometimes refer to these two particles
as ``spectators'', for reasons that will become clear below.
In infinite volume $\vec k$ and $\vec k'$ are continuous, but
the quantization condition, Eq.~(\ref{eq:qcond}),
depends only on $\Kdf$ for finite-volume momenta
satisfying $\vec k, \vec k' \in (2 \pi/L) \mathbb Z^3$. 
With $\vec k$, $\vec k'$ specified, the total momentum and
energy of the remaining two particles is also determined, 
separately for the in- and out-states. 
Thus the only remaining degrees of freedom are the incoming
and outgoing two-particle orbital angular momenta in their 
respective center-of-mass (CM) frames. 
We specify these using spherical harmonic indices:
$\ell, m$ for the in-state and $\ell',m'$ for the out-state. 
Altogether, for fixed $E$ (and $\vec P=0$),
$\Kdf$ depends on $\vec k', \ell', m'$ and $\vec k, \ell, m$. 
Since these quantities take discrete values,
it is convenient to view $\Kdf$ as a matrix,
i.e. $\Kdf = \Kdfin{k',\ell',m';k,\ell,m}$.
Equivalently, $\Kdf$ is a linear operator acting on a space
with orthonormal basis vectors $|\vec k,\ell,m\rangle$,
such that
\begin{equation}
\langle \vec k',\ell',m' | \Kdf | \vec k, \ell,m\rangle=\Kdfin{k',\ell',m';k,\ell,m}\,.
\end{equation}
 
The other factor in Eq.~(\ref{eq:qcond}), $F_3$, 
is a matrix acting on the same space. 
It is given by
\begin{equation}
F_{3} = \frac1{L^3} \frac{1}{2 \omega} 
\left [\frac{F}3 - F \frac{1}{{\K}^{-1} + F + G} F \right]\,,
\label{eq:Fthreealt}
\end{equation}
where we have used the form of the result given (up to trivial
rearrangements) in Appendix C of Ref.~\cite{us}.
Four new matrices enter Eq.~(\ref{eq:Fthreealt}):
$1/(2 \omega)$, $F$, $G$, and $\K$.
The first three are kinematical quantities, 
and will be described below.
We first discuss $\K$, which is given by
\begin{equation}
\Kin {k', \ell', m'; k, \ell, m} = 
\delta_{k'k} \delta_{\ell'\ell} \delta_{m'm} 
\frac{16 \pi E_{2,k}^*}{q_k^*} \tan \delta_{\ell}(q_k^*) 
\,.
\label{eq:Ktwodef}
\end{equation}
The physical interpretation of this infinite-volume quantity is that it describes a
process in which the spectator particles do not interact
(so that $\vec k=\vec k'$) while the other two particles scatter
(so that the two-particle CM angular momentum is conserved).
In the two-particle CM frame, the momentum of each particle is denoted
$q_k^*$, while their combined energy is $E^*_{2,k}$.
These are given, respectively, by
\begin{equation}
q^{*2}_k ={ E_{2,k}^{*2}/4-m^2} \ \ \ {\rm and}\ \ \
E^{*2}_{2,k} = {(E-\omega_k)^2 - \vec k^{\,2}}\,,
\label{eq:qkstardef}
\end{equation} 
where $\omega_k=\sqrt{\vec k^{\,2} +m^2}$.
Stripping away the Kronecker deltas from Eq.~(\ref{eq:Ktwodef}),
what remains is the two-particle K matrix, given in terms of
the physical, infinite-volume scattering phase shift $\delta_{\ell}(q_k^*)$.

As written, Eq.~(\ref{eq:Ktwodef}) is only valid above threshold,
i.e.~for $(q_k^{*})^2>0$. However, our formalism also requires 
$\K$ below threshold. This is because, as $\vec k^{\,2}$ increases,
$E^*_{2,k}$ drops below $2m$ and thus $(q^*_k)^2$ becomes negative.
The subthreshold result is defined in Ref.~\cite{us},
and is obtained from the above threshold result, (\ref{eq:Ktwodef}),
by two changes.
First, one analytically continues the scattering phase shifts
below threshold in the standard way using threshold expansions.
For example, for $\ell=0$, one uses
\begin{equation}
q^{-1} [\tan \delta_0(q)] = -a \left [1 + \frac{1}{2} r a q^2 +
  \mathcal O[(a q)^4] \right] \,,
\label{eq:psexp}
\end{equation}
which is valid for both positive and negative $q^2$.
Here $a$ is the scattering length in the nuclear physics
convention,\footnote{%
{The convention is such that $a>0$ for repulsive two-body interactions and $a<0$ for attractive. Thus we expect the proportionality factor in $\DEth \propto a/L^3 + \mathcal O(1/L^4)$ to be positive. [See the text after Eq.~(\ref{eq:DeltaEdef}) above.]}} 
and $r$ is the effective range. Similar expansions
exist for the higher partial waves, but we will only need the result 
\begin{equation}
q^{-1} [\tan \delta_\ell(q)] = \mathcal O(q^{2 \ell}) \,.
\label{eq:K2ell}
\end{equation}
In addition to the analytic continuation of the phase shift, the
subthreshold definition of $\K$ includes a term related to
the two-particle unitary cusp (and involving the cutoff function $H$ introduced below).
However, this term does not contribute to any power
of $1/L$ when doing an expansion about the threshold energy. 
We thus do not describe it in this work.

We now define the remaining matrices contained in $F_3$.
The first is a simple diagonal kinematical matrix,
\begin{equation}
\label{eq:omegamatdef}
\left[ \frac{1}{2 \omega} \right]_{k',\ell',m';k,\ell,m}  \equiv
\delta_{k'k} \delta_{\ell'\ell} \delta_{m'm} \frac{1}{2 \omega_k}\,.
\end{equation}
The second, $G$, resembles the three-particle nonrelativistic propagator,
decorated by angular dependence. It has both diagonal and off-diagonal
entries:
\begin{equation}
\label{eq:Gdef}
G_{p, \ell', m' ; k, \ell, m} 
\equiv
\left(\frac{k^*}{q_p^*}\right)^{\ell'} 
\frac{4 \pi Y_{\ell',m'}(\hat  k^*) 
H(\vec p\,) H(\vec k\,) Y_{\ell,m}^*(\hat p^*)} 
{2 \omega_{kp} (E - \omega_k - \omega_p - \omega_{kp})}
\left(\frac{p^*}{q_k^*}\right)^\ell 
\frac{1}{2 \omega_k L^3} \,.
\end{equation}
Here $q_p^*$ is defined as for $q_k^*$ in Eq.~(\ref{eq:qkstardef})
except with $k \to p$,
$\vec k^*$ is the result of boosting the vector $\vec k$ 
with velocity $\vec \beta_p = \vec p/(E - \omega_p)$,
and $\vec p^*$ is defined by a similar boost with $\vec k\leftrightarrow \vec p$.
In addition,
$\omega_{kp}=\sqrt{(\vec k+\vec p \,)^2 +m^2}$ is the on shell energy
of the particle with the ``third'' momentum coordinate, $-\vec k-\vec p$.
Finally, $H$ is a cutoff function, defined by\footnote{%
Other choices of the function $J$ are possible, as discussed in
Ref.~\cite{us}, but this is the form we use for numerical evaluations.}
\begin{equation}
H(\vec k) = J([E^*_{2,k}/(2m)]^2) \,,\quad
J(x) \equiv
\begin{cases}
0 \,, & x \le 0 \,; 
\\ 
\exp \left( - \frac{1}{x} \exp \left [-\frac{1}{1-x} \right] \right ) \,, 
& 0<x \le 1 \,; 
\\ 
1 \,, & 1<x \,.
\end{cases}
\label{eq:HJdef}
\end{equation}
It ensures that the boosts needed to obtain $\vec p^*$ and $\vec k^*$
are well defined. A key property of $J(x)$ is that it is smooth.
In particular, since $J(x)=1$ for $x\ge 1$, all its
derivatives vanish as $x\to 1^-$. Thus the function remains unity
to all orders in a Taylor expansion about $x=1$.
For further discussion of $J$ and $H$ see Ref.~\cite{us}.

The last matrix, $F$, is a generalization of the zeta functions
introduced in Ref.~\cite{Luscher:1986n2}:
\begin{align}
F_{k', \ell',m';k,\ell,m} 
& \equiv 
\delta_{k'k} F_{\ell',m';\ell,m}(\vec k)\,,
\label{eq:Fdef1}
\\
F_{\ell',m';\ell,m}(\vec k)
&=
F^{i\epsilon}_{\ell',m';\ell,m}(\vec k)
+
\rho_{\ell', m'; \ell, m}(\vec k) \,,
\label{eq:Fdef2}
\\
F^{i\epsilon}_{\ell',m';\ell,m}(\vec k)
&=
\frac12 \left[\frac{1}{L^3} \sum_{\vec a} - \int_{\vec a} \right] 
\frac{{4 \pi} Y_{\ell',m'}(\hat a^*) Y_{\ell,m}^*(\hat a^*) H(\vec k)
  H(\vec a\,)H(\vec b_{ka})}
{2 \omega_a 2 \omega_{ka}
(E - \omega_k - \omega_a -  \omega_{ka} + i \epsilon)}
\left(\frac{a^*}{q_k^*}\right)^{\ell+\ell'} \,.
\label{eq:Fdef3}
\end{align}
Here $\int_{\vec a} \equiv \int d^3 a/(2 \pi)^3$, while the sum over
$\vec a$ runs over all finite-volume momenta. 
$\vec a^*$ is the vector obtained by boosting $\vec a$ to the
two-particle CM frame, treating $\vec k$ as the spectator momentum, 
i.e.~boosting with velocity $\vec \beta_k = \vec k/(E - \omega_k)$.
Finally, $\rho$ is a phase space factor defined by
\begin{align}
\label{eq:rhodef}
\rho_{\ell',m';\ell,m}(\vec k)& \equiv 
\delta_{\ell'\ell} \delta_{m'm} H(\vec k) \tilde\rho(E_{2,k}^*)\,,
\\
\tilde\rho(E_{2,k}^*) &\equiv \frac{1}{16 \pi {E^*_{2,k}}} \times
\begin{cases} 
-  i q_k^* & (2m)^2< E_{2,k}^{*2} \,, 
\\ 
\vert q_k^* \vert &   0<E_{2,k}^{*2} \leq (2m)^2 \,,
\end{cases}
\label{eq:rhotildedef}
\end{align}
The addition of the $\rho$ term to $F^{i\epsilon}$ in Eq.~(\ref{eq:Fdef2})
changes the pole prescription from $i\epsilon$ to the 
``$\widetilde{\rm PV}$'' prescription defined in Ref.~\cite{us}.

We close this section by rearranging the matrices appearing 
in the quantization condition in two minor ways.
The first takes care of
the powers of $1/q_p^*$ or $1/q_k^*$ 
(which we collectively refer to as $1/q^*$) contained in $G$ and $F$.
Since we will find that $q^*\sim 1/L$,
these terms apparently lead to positive powers of $L$, 
complicating the development of the threshold expansion. 
These powers of $1/q^*$ are, however, misleading, since they are canceled by corresponding
positive powers contained within $\K$ and $\Kdf$. This is shown
for $\K$ by the result (\ref{eq:K2ell}), and for $\Kdf$ by a general
result shown in Appendix A of Ref.~\cite{us}.
It is thus preferable to make this cancellation explicit
by introducing factors of the matrix
\begin{equation}
Q_{k',\ell',m';k,\ell,m} \equiv \delta_{k'k}\delta_{\ell'\ell}\delta_{m'm}
(q_k^*)^\ell\,.
\label{eq:Qnewdef}
\end{equation}
The second change is to insert factors of the matrix $1/(2\omega)$
and its inverse $(2\omega)$ such that the symmetric matrix
$(2\omega)^{-1} G$ appears.

Specifically, we introduce
\begin{equation}
\tilde F_3 = Q F_3 Q\,,\ \ 
\tKdf = Q^{-1} \Kdf Q^{-1}\!, \ \
\tK = (2\omega) Q^{-1} \K Q^{-1}\!,\ \
\tilde F = (2\omega)^{-1} Q F Q\,,\ {\rm and}\ 
\tilde G = (2\omega)^{-1} Q G Q\,,
\label{eq:tildedefs}
\end{equation}
in terms of which the quantization condition becomes
\begin{equation}
\det [1 + \tilde F_{3} \tKdf]  =  0 \,,
\label{eq:qcond2}
\end{equation}
where
\begin{equation}
 \tilde F_{3} = \frac1{L^3} 
\left [\frac{\tilde F}3 - \tilde F 
\frac{1}{\cal H}
\tilde F \right] \,,
\label{eq:F3new}
\end{equation}
with
\begin{equation}
{\cal H} \equiv {\tK}^{-1} + \tilde F + \tilde G\,.
\label{eq:Hdef}
\end{equation}
We stress that both $\tKdf$ and 
$\tK$ have a well defined limit as
$q^*\to 0$, and indeed are functions of $(q^*)^2$ that can be
analytically continued to negative values.
We also note that $\tilde G$, $\tilde F$ and $\tK$, and thus also
${\cal H}$, are hermitian.


\section{Threshold expansion}
\label{sec:thr}

To develop the $1/L$ expansion we need to know how the various
quantities entering the quantization condition, Eq.~(\ref{eq:qcond}), scale 
with $1/L$ when $E\approx 3m$. Specifically, recalling that
$\vec k=2\pi \vec n/L$ is one of the matrix indices on the
quantities in (\ref{eq:qcond}),
we can work out the scaling assuming that $n=|\vec n| = {\cal O}(L^0)$ so that 
$k = {\cal O}(1/L) \ll m$. 
This is the same as assuming that important contributions to
the sums over matrix indices occur when all three
particles are nonrelativistic.
This assumption is naive, since the sums actually range up to values
of $\vec k$ where $H(\vec k)=0$, for which $k\sim m$. It turns out that the naive scaling gives the correct prediction for the first three orders in the $1/L$ expansion of $\DEth$. We demonstrate this in Sec.~\ref{sec:solveDeltaE}, where we also show how to reach the correct result for the $1/L^6$ contribution, for which the naive scaling is insufficient.

As we explain in detail in the first subsection below, the assumption $|\vec n| = {\cal O}(L^0)$, together with the assumed form
(\ref{eq:DeltaEexp}) for $\DEth$, allows one to determine the scaling with $1/L$ of
each of the components of the matrices entering into the quantization
condition. We find that the elements of $\tKdf$ are of ${\cal O}(L^0)$,
which is simply the statement that this is an infinite-volume quantity with a nonzero
limit at threshold.
The dominant contributions to $\tilde F$ and $\tilde G$ are also of ${\cal O}(L^0)$,
so that $\tilde F_3 \sim 1/L^3$ due to the explicit volume factor in Eq.~(\ref{eq:F3new}).
Naively, one might conclude that  $\tilde F_3 \tKdf \sim 1/L^3$ and cannot
cancel the contribution from the unit matrix in Eq.~(\ref{eq:qcond2}), as would be necessary
to satisfy the quantization condition.
There are two ways to avoid this conclusion. First, the determinant involves a product
over ${\cal O}(L^3)$ matrix indices, 
and this multiplicity factor can cancel the $1/L^3$ in $\tilde F_3$.
Second, the matrix $\mathcal H$ can, for an appropriately tuned energy, have an eigenvalue
of $\mathcal O(1/L^3)$, due to cancellations between the terms in Eq.~(\ref{eq:Hdef})
[which are each of ${\cal O}(L^0)$]. This leads to $\tilde F_3$ scaling as $\mathcal O(L^0)$.
Both mechanisms turn out to contribute in the solution to the quantization condition,
and we describe them in turn.

To illustrate the impact of having $\mathcal O(L^3)$ matrix indices, we expand
the determinant in terms of cofactors\footnote{%
$C_{k \ell m}$ is the determinant of the matrix reached by removing the $000$th row 
and the $k \ell m$th column from $1 +  \tilde F_{3} \tKdf $, multiplied by
an alternating phase. }
\begin{align}
\det [ 1 + \tilde F_{3} \tKdf ] & = \left(1 + [\tilde F_{3} \tKdf]_{000;000}\right)C_{000} 
+ \sum_{\{k  \ell  m\} \neq 0} [ \tilde F_{3} \tKdf]_{000;k \ell m} C_{k \ell m}\,.
\label{eq:expanddet}
\end{align}
We focus on the second term.
From the discussion above, we know that the matrix elements
$[ \tilde F_{3} \tKdf]_{000;k \ell m}$ scale as $1/L^3$. 
Now we use the result that the infinite-volume limit of $(1/L^3)\sum_{\vec k}$
acting on a smooth function equals the integral, $\int d^3k/(2\pi)^3$, of that function.
Assuming that $C_{klm}$ scales as $L^0$, this implies that the second term
in (\ref{eq:expanddet}) in fact scales as $L^0$ rather than as $1/L^3$.
To determine the actual scaling of $C_{klm}$, one would
need to iteratively repeat the cofactor analysis, removing increasingly more rows and columns and evaluating determinants.
It is plausible that this could lead to additional $L^3$ enhancements. 
In this study, however, we are able to avoid this complicated line of analysis, 
by recasting the
quantization condition in a form that, for studying the threshold energy, is simpler to
handle. 
We thus use Eq.~(\ref{eq:expanddet}) only to emphasize that the naive scaling of terms can be invalidated by
the presence of sums over the $\mathcal O(L^3)$ indices, leading to
a potential proliferation of contributions. This observation will play a central
role in the subsequent analysis.

To illustrate the second mechanism needed to find the threshold solution of the quantization condition,
we adopt the naive scaling worked out in the next subsection.
In this scaling, the dominant parts of $\tilde F$ and $\tilde G$ are, respectively,
$\tilde F_{00}\equiv \tilde F_{000;000}$ and 
$\tilde G_{00}\equiv \tilde G_{000;000}$, 
both of which scale as $L^0$ (as do all elements of $\tKdf$).
Here we are introducing the abbreviation that the subscript $00$ refers
to the matrix element with $\vec k=\vec k'=\vec 0$ and $\ell=\ell'=m=m'=0$.
The dominant part of $\tilde F_3$ is then
\begin{equation}
\tilde F_{3;00}\equiv
\tilde F_{3;000,000} \approx - \frac1{L^3} 
\tilde F_{00} [\mathcal H^{-1}]_{00}\tilde F_{00} \,,
\end{equation}
with all other matrix elements suppressed by additional powers of $1/L$.
If this were the entire story, the quantization condition would collapse,
as $L\to \infty$, to the algebraic equation
\begin{equation}
1 + \tilde F_{3;00} \tKdfin{00} = 0\,.
\label{eq:qcondnaive}
\end{equation}
This equation can be solved if $\Delta E$ [of the form shown in
Eq.~(\ref{eq:DeltaEexp})] can be tuned such that
$\mathcal H$ has an eigenvalue that behaves as $c/L^3$.
We call this putative small eigenvalue $\lambda_0$.
It is also necessary that the corresponding
eigenvector, $|\lambda_0\rangle$, have nonzero overlap
with $|\vec 0, 0,0\rangle$ when $L\to\infty$. In that case 
$[\mathcal H^{-1}]_{00}\sim L^3$, so that $\tilde F_{3;00} \sim L^0$ and
the quantization condition (\ref{eq:qcondnaive}) can be satisfied if $\Delta E$ is
tuned so that the constant $c$ has the appropriate value.
The requisite tuning of the eigenvalue of $\mathcal H$ is possible because,
as can be seen from Eq.~(\ref{eq:Hdef}), $\mathcal H_{00}$ consists
of three terms of $\mathcal O(L^0)$, two of which ($\tilde F$ and $\tilde G$)
depend on $\Delta E$ (as shown in the next subsection).

To obtain the correct expression for the energy of the near-threshold state one
must combine the two mechanisms. The first mechanism alone would
require a cancellation between quantities in which all finite-volume
sums have been replaced by integrals, so that dependence on $L$ is lost.
This cannot lead to the desired volume dependence of Eq.~(\ref{eq:DeltaEexp}).
The second mechanism does lead to such a volume dependence---indeed,
as we show below, in order that $\lambda_0\sim 1/L^3$ we must remove
$L^0$, $1/L$ and $1/L^2$ contributions from $\lambda_0$, and this fixes the coefficients
$a_3$, $a_4$ and $a_5$ in $\DEth$. However, to determine the
$a_6/L^6$ term in $\DEth$, it turns out that we must control an infinite
number of contributions arising because of the first mechanism.

As noted above, we have not found it fruitful to work directly with the expansion 
given in Eq.~(\ref{eq:expanddet}). Instead, after some trial and error, we have found
that an alternative form of the quantization condition allows a simpler analysis.
This is
\begin{equation}
\label{eq:strongqcond}
\lim_{E \rightarrow 3 m + \DEth} \langle \lambda_0 \vert \tilde F  \tKdf 
\frac{1}{1 \! + \! \tilde F_3 \tKdf } \tilde F \vert \lambda_0 \rangle   = \infty \,,
\end{equation}
where $\vert \lambda_0 \rangle$ is the eigenvector of $ \mathcal H$ introduced
above whose eigenvalue, $\lambda_0$, is tuned to be of $\mathcal O(1/L^3)$. 
We will provide motivation for this form shortly, but first explain why it is valid.
We begin by noting that we expect there to be only one eigenvalue that can be tuned in this
way, since only in the element $\mathcal H_{00}$ can the requisite
cancellation occur. This is consistent with our expectation that there is only a single
near-threshold state.
Next we note that the matrix element in Eq.~(\ref{eq:strongqcond})
can diverge if $\tilde F$ diverges or if one of the eigenvalues of
$1+\!\tilde F_3 \tKdf$ vanishes.\footnote{%
Divergences in eigenvalues of $\Kdf$ do not give a solution 
as they cancel between numerator and denominator.} 
The divergence of $\tilde F$ only occurs at non-interacting energies, 
and thus does not lead to interesting solutions.
We avoid them by requiring $\DEth$ to have the form indicated in
Eq.~(\ref{eq:DeltaEexp}), which differs from all non-interacting energies
once $L$ is large enough.
With this proviso we see that, whenever Eq.~(\ref{eq:strongqcond}) holds, 
the original quantization condition, Eq.~(\ref{eq:qcond2}), is also
satisfied.  In fact, Eq.~(\ref{eq:strongqcond}) is a stronger
condition than (\ref{eq:qcond2}) because it requires that the
eigenvector of $1+\!\tilde F_3 \tKdf$ corresponding to the vanishing
eigenvalue has nonzero overlap with the vectors $\tilde F \vert
\lambda_0 \rangle$ and $\tKdf \tilde F \vert \lambda_0 \rangle$.

A more physical motivation for the condition (\ref{eq:strongqcond}) is that it corresponds
approximately to finding the pole in the  correlation function 
\begin{equation}
C_{\phi^3}(E) = \int d\tau\, e^{i (iE) \tau} \langle \tilde \phi(\tau,\vec 0)^3 \ \tilde\phi(0,\vec 0)^3 \rangle\,,
\end{equation}
with $\tilde \phi(\tau, \vec k)$ the spatial Fourier transform, in the finite box, of
a scalar field coupling to a single particle. 
Here it is understood that the $\tau$ integral is performed for real $iE$. 
The resulting function can then be analytically continued into the entire 
complex $E$ plane, with the energy poles then appearing on the real $E$ axis.
This correspondence holds because (as shown below) $|\lambda_0\rangle$ differs from the
free particle state $|\vec 0,0,0\rangle$ by factors that vanish as $L\to \infty$.
In addition, the quantity $\tKdf (1 + \tilde F_3 \tKdf)^{-1}$ expands to
a geometric series in which, following the analysis of Ref.~\cite{KtoM}, 
we can think of $\tKdf$ as a local three-particle interaction, 
while the intervening  factors of $\tilde F_3$ incorporate all possible two-to-two
scatterings in finite volume.
The form of the correlator $C_{\phi^3}(E)$ is such that, if one were to use it in a numerical lattice calculation, one would pick out the near-threshold state. This is the case because the deviation of
the true state from the noninteracting state falls as a power of $1/L$

In any case, what matters in the following is that Eq.~(\ref{eq:strongqcond}) is a valid
form for the quantization condition. To see its utility, we define
\begin{align}
\tilde F_3 & \equiv \overline F_3 + F^{\lambda_0}_3 \,,
\label{eq:Fbardef}\\
 F^{\lambda_0}_3 & \equiv  -   \tilde F \vert \lambda_0 \rangle \frac{1}{\mathcal N_0    L^3 \lambda_0} \langle { \lambda_0} \vert  \tilde F \,, 
\end{align}
where $\langle \lambda_0 \vert \lambda_0 \rangle = \mathcal N_0$.\footnote{%
We use an unnormalized state $|\lambda_0\rangle$ since this proves convenient when
studying this state and its eigenvalue using
Raleigh-Schr\"odinger perturbation theory, as we do in Sec.~\ref{sec:explambda}.}
In words, Eq.~(\ref{eq:Fbardef}) splits $\tilde F_3$ into a part arising from the small eigenvector
of $\mathcal H$ and the remainder $\overline F_3$, which is not enhanced when
$\Delta E$ is tuned.
Substituting this form into our new quantization condition and performing straightforward
manipulations, we find
\begin{align}
\langle \lambda_0 \vert \tilde F   \tKdf 
\frac{1}{1\!+\!  \tilde  F_{3}  \tKdf}  \tilde F \vert \lambda_0 \rangle
& = \langle \lambda_0 \vert \tilde F    
\frac{1}{[\tKdf]^{-1}\!+\!  \overline F_{3}  \!+\!  F^{\lambda_0}_{3} }  
\tilde F \vert \lambda_0 \rangle \,, \\[5pt]
& = {\cal Z} \frac1{1 - {\cal Z}/({\cal N}_0 L^3 \lambda_0)}\,,
\label{eq:MdfLqcond}
\end{align}
where
\begin{equation}
{\cal Z} = 
  \langle \lambda_0 \vert \tilde F   \tKdf \frac{1}{1\!+\!  \overline  F_{3}  \tKdf}  \tilde F \vert \lambda_0 
  \rangle \,.
\label{eq:Zdef}
\end{equation}
We now see the reason for placing factors of $\tilde F$ next to the external states
in the quantization condition (\ref{eq:strongqcond}).
This mirrors the factors that appear in $F_3^{00}$, and leads to a simple final expression
(\ref{eq:MdfLqcond}) involving only the matrix element ${\cal Z}$.

Using Eq.~(\ref{eq:MdfLqcond}) we see that the quantization condition can
be rewritten as
\begin{equation}
{\cal Z} = {\cal N}_0 L^3 \lambda_0\,.
\label{eq:qcondnomat}
\end{equation}
We stress that although ${\cal Z}$ has a very similar form to the quantity
appearing in the quantization condition (\ref{eq:strongqcond}), it does not diverge near threshold.
This is because the enhanced contribution to $\tilde F_3$ has been removed,
and $\overline F_3$ is of ${\cal O}(1/L^3)$ for all near-threshold energies.
In fact, as we show below, ${\cal Z}$ is related to the 
divergence-free three-particle amplitude 
at threshold, $\Mdf$.

To use Eq.~(\ref{eq:qcondnomat}) we tune the coefficients
$a_3$, $a_4$ and $a_5$ in $\DEth$ such that 
$\lambda_0\sim 1/L^3$. Then we fix $a_6$ by enforcing (\ref{eq:qcondnomat}).
Clearly this form of the quantization condition is much simpler than
the original version, Eq.~(\ref{eq:qcond}), 
since it no longer requires evaluating the formally infinite-dimensional determinant. 
This simplicity comes, however, at a cost in generality---our new form is only
useful for studying the near-threshold state. 

In the following subsections we use this reduced quantization condition 
to determine the $1/L$ expansion of the threshold energy shift. 
This analysis is organized as follows. 
We begin in the following subsection by determining the $1/L$ scaling properties of 
$\tK$, $\tilde G$ and $\tilde F$. Next, in Sec.~\ref{sec:explambda}, 
we use these inputs to develop the perturbative expansion of $\lambda_0$ and the
corresponding state $|\lambda_0\rangle$. 
Following this, in Sec.~\ref{sec:MbartoMdf} we prove an important identity 
relating a matrix element entering the quantization condition and the infinite-volume
divergence-free three-particle scattering amplitude.
We manipulate this result further in Sec.~\ref{sec:MdftoMthr}, 
to reach our final threshold three-particle observable, denoted $\Mthr$. 
Finally in Sec.~\ref{sec:solveDeltaE} we combine results to expand 
Eq.~(\ref{eq:qcondnomat}) in powers of $1/L$ and determine the coefficients in $\DEth$.

\subsection{Scaling of matrix components with $1/L$}
\label{sec:KFGscaling}

In this subsection we determine how the elements of
the matrices $\tK$, $\tilde F$ and $\tilde G$ scale with $1/L$
in the regime where the spectator-momentum matrix index satisfies $k\sim 1/L \ll m$. 
We assume that $\Delta E$ scales as $1/L^3$ throughout.

We repeatedly use several simple kinematic results that 
follow from the definitions in Eq.~(\ref{eq:qkstardef}).
In the special case $\vec k=0$ we have the exact results
\begin{equation}
\omega_k=\omega_0=m\,,\quad 
q_0^{*2} = m \Delta E + \frac{\Delta E^2}{4}\equiv q^2 \,, \ \ {\rm and} \ \
E^*_{2,0}= 2m + \Delta E = 2 \omega_q\,,
\label{eq:E20q0}
\end{equation}
where we have introduced the convenient abbreviation $q$ for the three-momentum
of each of the non-spectator particles in the case that the spectator has zero momentum.
We note that $q^2\sim \Delta E\sim 1/L^3$.
For general $\vec k=2\pi \vec n/L \neq \vec 0$, with $n\sim {\cal O}(1)$,
we expand in powers of $1/L$, finding
\begin{align}
\omega_k &= m \left(1 + \frac{k^2}{2m^2} + {\cal O}[(mL)^{-4}] \right) \,,
\label{eq:omegakexp}
\\
E_{2,k}^* &= 2m \left(1 - \frac{3 k^2}{8 m^2} + \frac{\Delta E}{2m}
+ {\cal O}[(mL)^{-4}] \right) \,,
\label{eq:E2kstarexp}
\\
q_k^{*2} & = -\frac{3 k^2}{4} + m \Delta E + m^2 \mathcal O[(mL)^{-4}] \,.
\label{eq:qstarexp}
\end{align}
Note that, unlike for $\vec k=0$, in this case the CM frame of the non-spectator pair is
moving relative to the rest frame of the finite volume.

We consider first the $1/L$ scaling of $\tK$, which we recall is a diagonal matrix.
Since this is an infinite-volume quantity, $L$ dependence enters only through $\Delta E$. 
The leading term is of ${\cal O}(L^0)$ and is simply given by the value of $\tK$ at
threshold in the appropriate partial wave. As noted above,
this is non-vanishing for all $\ell,m$ because of the factors of $Q^{-1}$ in the definition
(\ref{eq:tildedefs}).
It turns out that the only explicit expression we need is for the
$\vec k=0, \ell=0, m=0$ element.\footnote{%
Higher partial waves are suppressed because, in the matrix products that arise,
these are always multiplied by entries of $\tilde G$ or $\tilde F$ with $\ell \neq 0$.}
This can be obtained by inserting the threshold expansion, Eq.~(\ref{eq:psexp}), 
into the definitions (\ref{eq:Ktwodef}) and (\ref{eq:tildedefs}),
and using the kinematic results of Eq.~(\ref{eq:E20q0}):
\begin{align}
\tKin {00}&=
-64 \pi m^2 a \left\{1 +
\frac{\Delta E}{2 m} [1 + r a m^2]  + 
\mathcal O\left[\frac1{(mL)^{6}}\right] \right\}\,.
\label{eq:K00res}
\end{align}

At this stage, we reiterate that we are considering in this subsection only what we have
called the ``naive'' scaling behavior, valid when $k \sim 1/L \ll m$.
It turns out that {\em all} entries of $\tK$ (i.e. all $\vec k$, $\ell$ and $m$) 
actually contribute to $\Delta E$ at $\mathcal O(1/L^6)$,
due to the high-momentum ends of the sums over indices.
This is explained in Sec.~\ref{sec:explambda}. 

The scaling of the elements of $\tilde G$ is more complicated.
Recall that $\tilde G$ is given by Eq.~(\ref{eq:Gdef}) with the 
factors of $q_{k}^*$ removed and an overall factor of $1/(2\omega_p)$ included:
\begin{equation}
\label{eq:Gtildef}
\tilde G_{p, \ell', m' ; k, \ell, m} 
\equiv
\frac{1}{L^3} \frac{1}{2 \omega_p} \frac{4 \pi (k^*)^{\ell'}   Y_{\ell',m'}(\hat  k^*) 
H(\vec p\,) H(\vec k\,) (p^*)^\ell  Y_{\ell,m}^*(\hat p^*)} 
{2 \omega_{kp} (E - \omega_k - \omega_p - \omega_{kp})}
\frac{1}{2 \omega_k } \,.
\end{equation}
We begin with the generic case in which  one or both of $\vec k$ and $\vec p$
are of ${\cal O}(1/L)$, from which it follows that both
$\vec p^{\,*}$ and $\vec k^*$ are also of this order.
Noting that the energy denominator then behaves
as $E-\omega_k-\omega_p-\omega_{kp}\sim 1/L^2$, we see from Eq.~(\ref{eq:Gtildef}) that the generic scaling is
\begin{equation}
\tilde G_{p,\ell',m';k,\ell,m} \sim \frac1{L^{1+\ell+\ell'}}
\qquad
(\vec k\ne0 \ {\rm and/or}\ \vec p\ne0).
\label{eq:tGgenericscaling}
\end{equation}
The exceptions to this scaling are the $\vec k=\vec p=0$ elements.
These are special because $k^*$ and $p^*$ now vanish, 
and the energy denominator scales as $E-3m=\Delta E\sim 1/L^3$ rather than $1/L^2$.
These results imply that the $\vec k=\vec p=0$ elements of $\tilde G$ vanish unless $\ell=\ell'=0$,
while the $00$ element is of ${\cal O}(L^0)$,
rather than ${\cal O}(1/L)$ as the generic scaling would predict.

The upshot is that, in the naive scaling regime, the dominant
contributions are from the $\ell=\ell'=0$ entries of $\tilde G$. 
These can be obtained directly from the definition Eq.~(\ref{eq:Gtildef}). 
We quote here only the $\vec k=\vec p=0$ component
\begin{equation}
\tilde G_{00} = \frac1{8 m^3 \Delta E L^3}\,.
\label{eq:G00res}
\end{equation}
As we show in Sec.~\ref{sec:explambda}, it turns out that only the $\ell=\ell'=0$ entries of $\tilde G$ contribute to $\DEth$ through $\mathcal O(1/L^5)$.
As for $\tK$, all entries of $\tilde G$ contribute to $\DEth$ at $\mathcal O(1/L^6)$ 
due to the high-momentum ends of the sums.

Finally, we describe the scaling of the elements of $\tilde F$,
which we recall is given by Eqs.~(\ref{eq:Fdef1})-(\ref{eq:Fdef3})
multiplied by $(2\omega_k)^{-1}(q_k^*)^{\ell+\ell'}$:
\begin{equation}
\tilde F_{k', \ell',m';k,\ell,m} 
\equiv 
\delta_{k'k} \frac12 \left[\frac{1}{L^3} \sum_{\vec a} - \int_{\vec a} \right] 
\frac{{4 \pi} (a^*)^{\ell+\ell'} Y_{\ell',m'}(\hat a^*) Y_{\ell,m}^*(\hat a^*) H(\vec k)
  H(\vec a\,)H(\vec b_{ka})}
{2 \omega_k 2 \omega_a 2 \omega_{ka}
(E - \omega_k - \omega_a -  \omega_{ka} + i \epsilon)}
+ \delta_{k'k} \frac{(q^*_k)^{\ell+\ell'}}{2 \omega_k} \rho_{\ell', m'; \ell, m}(\vec k) \,.
\label{eq:Ftilexplicit}
\end{equation}
Note that $\tilde F$ diverges whenever $E$ equals the 
sum of the energies of three free particles, each having
a finite-volume momentum (and with the total momentum vanishing). 
The value of $E$ we are interested in---the near-threshold
energy level in the presence of interactions---avoids these divergences.
However, the fact that these poles lie nearby can enhance the scaling of $\tilde F$.

To determine the nature of this enhancement,
we rewrite $\tilde F$ in terms of dimensionless variables,
using manipulations mirroring those used in Ref.~\cite{Kim:2005}. 
Dropping
contributions to the summand of the sum-integral difference that are nonsingular 
(and which thus lead only to exponentially suppressed contributions to $\tilde F$), 
we find 
\begin{align}
\tilde F_{k',\ell',m';k,\ell,m}
&= \delta_{k'k} \left\{
\left(\frac{H(\vec k)}{16 \pi^2 \omega_k (E-\omega_k)}\right)
\left(\frac{2\pi}L\right)^{1+\ell+\ell'}
{\cal Z}_{\ell',m';\ell,m}(x^2,\vec n_k)
+ \frac{(q_k^*)^{\ell+\ell'}}{2\omega_k} \rho_{\ell',m';\ell,m}
\right\}\,,
\label{eq:Ftozeta}
\\
{\cal Z}_{\ell',m';\ell,m}(x^2,\vec n_k) &=
\left[\sum_{\vec n_a}-\int_{\vec n_a}\right]
\frac{r^{\ell'+\ell}\;Y_{\ell',m'}(\hat r)Y^*_{\ell,m}(\hat r)
H(\vec a)H(\vec b_{ka})}
{x^2-r^2 +i\epsilon}
\,,
\label{eq:zetadef}
\end{align}
where $x=q_k^* L/(2\pi)$, $\vec a=2\pi \vec n_a/L$,
and $\int_{\vec n_a}=\int d^3n_a$.
The vector $\vec r$ is related to $\vec n_a$ by
\begin{equation}
r_\parallel = \frac1\gamma\left(n_{a\parallel} - \vert \vec n_{k} \vert/2\right)\,,
\quad
r_\perp = n_{a\perp}\,,
\quad
\vec n_k = \frac{\vec k L}{2\pi}\,,
\quad
\gamma = \frac{E-\omega_k}{E_{2,k}^*}\,,
\label{eq:rdef}
\end{equation} 
where parallel and perpendicular are relative to the
momentum $-\vec k$ of the nonspectator pair.
Note that $r^2$ runs over all positive values and zero as $\vec n_a$ is varied.

The function in Eq.~(\ref{eq:zetadef}) is simply related
to the zeta functions defined in Ref.~\cite{Kari:1995}
(in a way described in Refs.~\cite{Kim:2005,Hansen:2012tf})
except that here we are using a different UV regularization.\footnote{%
Our functions are regulated by the product of $H$ functions,
whereas Ref.~\cite{Kari:1995} uses analytic regularization.}
The key property of this function for the present discussion is that, for fixed $\vec n_k$, as $L \to \infty$,
${\cal Z}_{\ell',m';\ell,m}(x^2,\vec n_k)$ limits to an $L$-independent function of $x^2$ that
is finite except for an infinite sequence of poles.
There is one pole for each term in the sum over $\vec n_a$, occurring
when $x^2$ equals the corresponding value of $r^2$.
These are exactly the poles mentioned above that occur
when $E=\omega_k+\omega_a+\omega_{ka}$.

We first consider $\vec k = \vec k' \neq \vec 0$, 
so that, using Eq.~(\ref{eq:qstarexp}), $x^2 = -3 n_k^2/4 + \mathcal O(1/L)$. 
Since $x^2$ is negative definite, it does not approach the poles of ${\cal Z}$,
which are all at $x^2\ge 0$. Thus there is no enhancement of the scaling,
and ${\cal Z}(x^2, \vec n_k) = {\cal O}(L^0)$.
The scaling of the first term in $\tilde F$ is therefore given by the explicit factors of $1/L$.
The $\rho$-dependent term has the same scaling, and so we conclude that
$\tilde F \sim 1/L^{1+\ell+\ell'}$ when $\vec k = \vec k' \neq \vec 0$. 

We next turn to $\vec k=\vec k'=\vec n_k=0$, in which case $x^2$ vanishes
in the infinite volume limit: $x^2=q^2L^2/(2\pi)^2 \sim 1/L$. Thus if ${\cal Z}$ has a pole
at $x^2=0$, there can be an enhancement in the scaling. For $\vec n_k=0$, Eq.~(\ref{eq:rdef}) gives $r^2=n_a^2$, 
so there is indeed pole at $x^2=0$, from the $\vec n_a=0$ term in the sum. 
However, this pole is present only when $\ell=\ell'=0$.
For nonvanishing angular momenta, the residue vanishes due to the
factor of $r^{\ell'+\ell}= n_a^{\ell+\ell'}$.
This implies that, even for $\vec k=\vec k'=0$, the scaling derived above
for $\vec k=\vec k'\ne 0$ 
holds when one or both of $\ell$ and $\ell'$ are nonvanishing.
The only special case is $\vec k=\vec k'=\ell=\ell'=0$.
Here the $\vec n_a=0$ term in the sum gives
$\mathcal Z(x^2,\vec 0) \sim 1/x^2 \sim L$, 
and thus $\tilde F_{00} \sim L^0$, i.e.~enhanced by one power of $L$ compared
to the generic scaling. In the subsequent analysis we will need the first four terms
in the $1/L$ expansion of $\tilde F_{00}$.
These are worked out in Appendix~\ref{app:tF00}, 
with the result
\begin{equation}
\tilde F_{00}  =   \frac{1}{16 m \omega_q} \left\{
\frac{1}{q^2 L^3} 
- \frac{{\cal I}}{4\pi^2 L}
- \frac{q^2 L^3 {\cal J}}{(4\pi^2 L)^2}
- \frac{(q^2 L^3)^2 {\cal K}}{(4\pi^2 L)^3}
+{\cal O}\left(\frac{1}{L^4}\right)\right\}\,.
\label{eq:tildeF00final}
\end{equation}
Here ${\cal I}$, ${\cal J}$ and ${\cal K}$ are numerical constants
defined in Appendix~\ref{app:tF00}. 

As with $\tK$ and $\tilde G$, 
the high-momentum entries of $\tilde F$ also contribute to $\Delta E$ 
at $\mathcal O(1/L^6)$. This contribution comes only from the $\rho$-dependent term, the second term in Eqs.~(\ref{eq:Ftilexplicit}) and (\ref{eq:Ftozeta}).

\subsection{Perturbative expansion of $\lambda_0$}

\label{sec:explambda}

In this subsection we develop the perturbative expansion of $\lambda_0$, the eigenvalue 
that appears on the left-hand side of our reduced quantization condition, Eq.~(\ref{eq:qcondnomat}). We recall that $\lambda_0$ is the eigenvalue $\mathcal H$
[defined in Eq.~(\ref{eq:Hdef})]
that can be tuned to be of $\mathcal O(1/L^3)$ by adjusting $\Delta E$. 
This tuning is required to satisfy Eq.~(\ref{eq:qcondnomat}). 
As already mentioned, $\mathcal H$ is generally $\mathcal O(L^0)$, so that $\Delta E$ must be adjusted to cancel three orders to achieve the desired scaling. 
Such a cancellation is only possible for the $00$ entry of $\mathcal H$, 
because only for this entry do $\tilde G$ and $\tilde F$ contain ${\cal O}(L^0)$ 
parts that can cancel with $\tK^{-1}$. It follows that $\lambda_0$ can be described 
as a perturbation of this entry and that the corresponding state, 
$\vert \lambda_0 \rangle$, is a perturbation of $|\vec 0, 0,0\rangle$. 

We now seek to determine an expression for $\lambda_0$ in terms of $\Delta E$. 
Since $\mathcal H$ is hermitian we can borrow technology from non-relativistic quantum mechanics. In particular, we analyze $\lambda_0$ using a method related to
Raleigh-Schr\"odinger perturbation theory (RSPT).
It proves convenient to first slightly rewrite our ``Hamiltonian'' in
terms of the two-particle scattering amplitude $\M$ instead of $\K$,
\begin{equation}
\mathcal H = {\tM}^{-1} + \tilde F^{i\epsilon} + \tilde G
\,,
\end{equation}
where 
\begin{align}
\M^{-1} &= \K^{-1} + \rho\,,
\label{eq:M2def}
\\
\tM^{-1} &= (2\omega)^{-1}Q {\M}^{-1} Q  
= \tK^{-1} + (2\omega)^{-1} Q \rho\, Q \,,
\label{eq:tildeM2def}
\\
\tilde F^{i \epsilon} &= (2 \omega)^{-1} Q F^{i \epsilon} Q 
=   \tilde F - (2\omega)^{-1} Q \rho \,Q\,,
\end{align}
and  $F^{i\epsilon}$ is defined in Eq.~(\ref{eq:Fdef3}).
The reason for this choice is that, for fixed $k \sim m$, $F^{i\epsilon}(\vec k)$
is exponentially suppressed as $L\to\infty$, since the summand of
the sum-integral difference in (\ref{eq:Fdef3}) is smooth.\footnote{%
For $\Delta E\sim 1/L^3$ and the spectator momentum $k\sim m$, the 
non-spectator pair are far below threshold, the energy denominator is
of $\mathcal O(m)$, and there are no poles.}
We use this result repeatedly in the following analysis.
The same is not true of $F(\vec k)$, due to the $\rho(\vec k)$ term in
Eq.~(\ref{eq:Fdef2}).

Next, we split $\mathcal H$ into a part $\mathcal H_0$ that contains all the
terms scaling as $L^0$ in the $k \sim 1/L$ regime, and the remainder, ${\cal H}_R$,
which is of ${\cal O}(1/L)$. As explained in the previous subsection, all nonzero
elements of $\tK$, as well as the components $\tilde F_{00}$ and $\tilde G_{00}$,
are of ${\cal O}(L^0)$. The $\rho$ terms are of ${\cal O}(1/L)$ and thus do not change
the scaling. Thus we introduce the subtracted quantities
\begin{align}
\slashed F^{i \epsilon} & \equiv \tilde F^{i \epsilon} 
- \vert \vec 0, 0, 0 \rangle \tilde F^{i \epsilon}_{00} \langle \vec0, 0, 0 \vert \,,
\label{eq:Fslashdef}
\\
\slashed G & \equiv \tilde G - \vert \vec 0,0,0 \rangle \tilde G_{00} \langle \vec 0, 0,0 \vert \,,
\label{eq:Gslashdef}
\end{align}
in which the $\mathcal O(L^0)$ component is excised, and split ${\cal H}$ as
\begin{align}
{\cal H}&={\cal H}_0 + {\cal H}_R\,,
\\
{\cal H}_0
&= 
{\tM}^{-1} 
+ \vert \vec 0, 0, 0\rangle (\tilde F^{i \epsilon}_{00}+\tilde G_{00}) \langle \vec 0,0,0 \vert  \,,
\label{eq:H0def}
\\
{\cal H}_R
&= \slashed F^{i \epsilon}+ \slashed G\,.
\label{eq:HRdef}
\end{align}
By construction, $\mathcal H_0$ is diagonal, with eigenvectors $|\vec k, l, m\rangle$,
and corresponding eigenvalues
\begin{align}
\lambda_0^{(0)} &\equiv \lambda_{000}^{(0)} =
\tilde {\mathcal M}^{-1}_{2;00} + \tilde F^{i\epsilon}_{00} + \tilde G_{00} 
=
\tilde {\mathcal K}^{-1}_{2;00} + \tilde F_{00} + \tilde G_{00} \,,
\\
\lambda_{klm}^{(0)} &= 
\frac{q_k^*}{16 \pi E_{2,k}^*} \cot\delta_\ell(q_k^*) + \tilde\rho(E_{2,k}^*)\,,
\qquad\qquad \{\vec k,l,m\} \ne \{\vec 0, 0, 0\}\,.
\end{align}
We see again that only $\lambda_0^{(0)}$ can be tuned to be small,
while all other eigenvalues are of $\mathcal O(L^0)$.
One subtlety in the following is that $\mathcal H_0$ is not necessarily hermitian,
since the eigenvalues with $\vec k=0$ but $\ell\ne 0$ can be complex.
This is because $\tilde\rho(E_{2,k}^*)$ 
[defined in Eq.~(\ref{eq:rhotildedef})] is imaginary if $\Delta E>0$,
which is possible if $\vec k=0$.
Nevertheless, the eigenvectors of $\mathcal H_0$ form an orthonormal basis, 
and this is sufficient for the subsequent analysis.
We note also that $\mathcal H_0$ does become hermitian when $\Delta E=0$,
i.e.~when $L\to\infty$. 

Using the results for $\tK$, $\tilde F_{00}$ and $\tilde G_{00}$ given in
Eqs.~(\ref{eq:K00res}), (\ref{eq:G00res}) and (\ref{eq:tildeF00final}), respectively,
as well as the kinematic relation (\ref{eq:E20q0}),
we can work out the $1/L$ expansion of $\lambda_0^{(0)}$.
We obtain
\begin{align}
\lambda_0^{(0)} &=
 -\frac{1}{64\pi m^2 a} \left[1- \frac{x(1+r a m^2)}{2 m^3 L^3} \right]
+ \frac1{16 m} \left[\frac1x
- \frac{{\cal I}}{4\pi^2 mL}
- \frac{{\cal J} x}{(4\pi^2 mL)^2}
-\frac{{\cal K} x^2}{(4\pi^2 mL)^3}
-\frac3{4 m^3 L^3}\right]
+ \frac1{8 m x}
+ {\cal O}\left(\frac1{L^4}\right)\,,
\label{eq:lambda0scaling}
\end{align}
where the first square bracket contains the expansion of $\tKin{00}$, the
second the expansion of $\tilde F_{00}$, the last term is
$\tilde G_{00}$, and we have introduced the dimensionless variable
\begin{equation}
x \equiv \Delta E L^3 m^2 \,,
\label{eq:xdef}
\end{equation}
which is of ${\cal O}(L^0)$.
In order to tune $\lambda_0$ to scale as $1/L^3$,
we will find that $\lambda_0^{(0)}$ itself must scale as $1/L^2$.
This is because the difference, $\lambda_0-\lambda_0^{(0)}$,
contains a term scaling as $1/L^2$ that must be canceled.
We defer details of the tuning to Sec.~\ref{sec:solveDeltaE},
except for one result. This concerns the cancellation
of the ${\cal O}(L^0)$ part of $\lambda_0^{(0)}$.
Using the result $2\tilde G_{00}=  \tilde F_{00} + {\cal O}(1/L)$
[which can be read off from Eq.~(\ref{eq:lambda0scaling})]
this cancellation requires $- 3 \tKin{00} \tilde F_{00} = 1 + {\cal O}(1/L)$.
We need this result in the following subsection.

We now work out the perturbative expansion for $\lambda_0$
and the corresponding eigenvector $|\lambda_0\rangle$ in powers of $\mathcal H_R$.
A standard starting point for developing RSPT is
\begin{align}
\vert \lambda_0\rangle &= \vert \lambda_0^{(0)}\rangle 
+ \Rzero 
(\mathcal H_R - \lambda_0 + \lambda_0^{(0)}) 
\vert \lambda_0\rangle\,,
\label{eq:RSPT}
\\
\Rzero &\equiv
 \frac{1-|\lambda_0^{(0)}\rangle\langle\lambda_0^{(0)}|}{\lambda_0^{(0)}-\mathcal H_0}
 \,,
 \label{eq:DeltaH0def}
\end{align}
where $\vert \lambda_0^{(0)}\rangle = \vert \vec 0, 0,0\rangle$ is the
unperturbed state. Note that, in this formulation,
$\vert\lambda_0\rangle$ satisfies
$\langle \lambda_0^{(0)}\vert \lambda_0\rangle=1$, implying
that $\vert \lambda^0 \rangle$ as defined in Eq.~(\ref{eq:RSPT}) is not normalized to unity,
${\cal N}_0=\langle \lambda_0\vert \lambda_0\rangle \ne 1$.
Iterating Eq.~(\ref{eq:RSPT}) yields
\begin{equation}
\vert \lambda_0 \rangle = \sum_{n=0}^\infty \vert \lambda^{(n)}_0 \rangle\,,
\end{equation}
with
\begin{equation}
\vert \lambda^{(n)}_0 \rangle \equiv 
\left[ \Rzero 
(\mathcal H_R - \lambda_0 + \lambda_0^{(0)}) \right]^n 
\vert \lambda^{(0)}_0 \rangle\,.
\label{eq:nthorderstate}
\end{equation}
Contracting with $\langle \lambda_0^{(0)}\vert \mathcal H$ leads to the
perturbative expansion for the eigenvalue
 \begin{align}
\lambda_{0} & = \lambda_{0}^{(0)}
+  
 \sum_{n=0}^\infty \lambda^{(n+1)}_{0}
\,, \\
\lambda^{(n+1)}_{0} & \equiv   
\left [  \mathcal H_R \bigg [ \Rzero
(\mathcal H_R - \lambda_{0} + \lambda_{0}^{(0)} )  \bigg ]^{n} \right ]_{00} 
\,.
\label{eq:PTforlambda}
\end{align}
To obtain standard RSPT one inserts the expansion for $\lambda_0$ and
reexpands in powers of $\mathcal H_R$.
We will not take this step but rather work with the forms above, containing
$\lambda_0$.
This is possible because we will find that, at the order in $1/L$ that we work,
we can set $\lambda_0^{(0)}-\lambda_0$ to zero on the right-hand sides of
Eqs.~(\ref{eq:nthorderstate}) and (\ref{eq:PTforlambda}).

We first analyze the perturbative shift to the eigenvalue.
Naively, since $\mathcal H_R\sim 1/L$, we might expect that a third-order
calculation is sufficient to obtain the desired accuracy, $\lambda_0 \sim 1/L^3$.
However, as described in the introduction to this section,
this scaling breaks down for $k\sim m$, and for such large momenta
it turns out that an all-orders summation is needed.

The first-order shift $\lambda_0^{(1)}$ vanishes,
since $\slashed G$ and $\slashed F^{i \epsilon}$ 
are both defined with vanishing $00$th component. 
Thus the first nonvanishing correction appears at second order:
\begin{equation}
\label{eq:lam2start}
\lambda^{(2)}_{0} = \left[ (\slashed F^{i \epsilon} + \slashed G)  
\Rzero (\slashed F^{i \epsilon} + \slashed G   ) \right ]_{00} \,.
\end{equation}
To obtain this form we have used the result
\begin{equation}
\Big [ (\slashed F^{i \epsilon} + \slashed G)  \Rzero 
(- \lambda_0 + \lambda_0^{(0)}) \Big ]_{00} = 0 \,,
\end{equation}
which follows from the fact that $\Rzero$ has all zeroes in its first column. 
We can further reduce $\lambda^{(2)}_0$ by using the fact that $\lambda_0^{(0)}$ 
will be tuned to be of $\mathcal O(1/L^2)$. This implies
\begin{equation}
\label{eq:dHexp}
 \Rzero = - \tM + \mathcal O(1/L^2) \,,
 \end{equation}
and substituting into Eq.~(\ref{eq:lam2start}) gives
\begin{equation}
 \label{eq:lam2explicit}
\lambda^{(2)}_{0} = \sum_{\vec k,\ell,m}  
\slashed G_{000;k \ell m} \left [- \tM + \mathcal O(1/L^2)  \right ]_{k\ell m; k \ell m}  
\slashed G_{k\ell m;000} \\
 + \sum_{\ell m}   
 \slashed F^{i \epsilon}_{000;0\ell m}  
 \left [ - \tM + \mathcal O(1/L^2)  \right ]_{0 \ell m; 0 \ell m}  
\slashed F^{i \epsilon}_{0 \ell m; 000}\,,
\end{equation}
where we have written out all sums explicitly. 
In writing this form we have used the facts that $\slashed F^{i\epsilon}$ is diagonal in $\vec k$,
that the slashed quantities have no $00$ element, 
and that $\slashed G_{000;0lm}$ vanishes whenever $\ell\ne 0$.

We want to pick out contributions falling no faster than $1/L^3$ from Eq.~(\ref{eq:lam2explicit}). 
We do so by keeping terms that have the desired scaling either in the low-momentum
($k\sim 1/L$) regime or in the high-momentum ($k\sim m$) regime, or both.
For low momenta, the dominant contribution comes from the first term with
$\ell=0$, for then $\slashed G={\cal O}(1/L)$. Thus the first term scales as $1/L^2$
(and the dominant contribution arises when intermediate angular momentum vanishes).
In the second term, only $\ell\ne 0$ contributes, with the leading term coming from
$\ell=4$. Since $\slashed F^{i \epsilon}_{000;040} = \mathcal O(1/L^5)$,
the second term scales as $1/L^{10}$, 
and can be dropped in the low-momentum regime.\footnote{%
This follows from the observation that $Y_{40}(\hat k)$ is the lowest spherical harmonic 
with $\ell\ne 0$
for which $(1/L^3) \sum_{\vec k} Y_{40}(\hat k) f(\vert \vec k \vert) \neq 0$, 
where $f(\vert \vec k \vert)$ is any radial function for which the sum converges.}
In fact, for this term this is the only relevant regime, since there is no sum over $\vec k$.

What remains is to analyze the first term in
Eq.~(\ref{eq:lam2explicit}) in the high-momentum regime, $k\sim m$.
Then the only explicit $L$ dependence arises from
the overall factor of $1/L^3$ in $\slashed G$.
At first sight this leads to a $1/L^6$ scaling since there are two factors of $\slashed G$.
However, the total number of terms in the high-momentum part of the sum scales as $L^3$,
canceling one of the factors of $1/L^3$.
This is just an application of the result that, for a smooth function\footnote{%
$\tM$ and $\slashed G$ are both smooth functions in the high-momentum
regime, since this corresponds (when $\Delta E\sim 1/L^3$) to the
far sub-threshold region.}
$f(\vec k)$,
\begin{equation}
\frac{1}{L^3} \sum_{\vec k} f(\vec k) = \int \! \frac{d^3k}{(2 \pi)^3} f(\vec k) 
+ \mathcal O(e^{- m L}) \,.
\end{equation}
The resulting integral is independent of $L$, and we are dropping 
exponentially suppressed corrections.
The conclusion is that the high-momentum contribution to $ \slashed G  [-\tM ]    \slashed G$ 
scales as $1/L^3$. 
While subleading to the low-momentum $1/L^2$ scaling, 
it is still of an order that we must keep. 
We also note that, in contrast to the low-momentum result, 
higher angular-momentum contributions are not suppressed when $k \sim m$.

The net result is that
\begin{equation}
\lambda^{(2)}_{0} = \left [ \slashed G  [-\tM ]    \slashed G \right ]_{00}   
+ \mathcal O \left( \frac{1}{L^4} \right ) \,,
\end{equation}
where no constraint is placed on the intermediate matrix indices.

We now turn to the third-order perturbative correction, which takes the form
\begin{align}
\lambda^{(3)}_{0} & =   
\left [   (\slashed F^{i \epsilon} + \slashed G ) \Rzero (\slashed F^{i \epsilon} + \slashed G - \lambda_{0} + \lambda_{0}^{(0)} ) \Rzero (\slashed F^{i \epsilon} + \slashed G  ) \right ]_{00} \,,\\
& =   
\left [     \slashed G  [- \tM ] (\slashed F^{i \epsilon} + \slashed G - \lambda_{0} + \lambda_{0}^{(0)} )[ - \tM  ]    \slashed G   \right ]_{00} + \mathcal O\left(\frac{1}{L^6} \right ) \,.
\label{eq:lam3two}
\end{align}
There are now two summed momenta, which we refer to as $k$ and $p$,
and to determine the scaling we must examine contributions from all possible momentum regimes.
First suppose both are of $\mathcal O(1/L)$, so that naive scaling can be applied.
Then the dominant contribution, scaling as $1/L^3$,
comes from the $s$-wave parts of each factor of $\slashed G$ and $\slashed F^{i \epsilon}$. 
This is the first example where $\slashed F^{i \epsilon}$ enters the result for $\lambda_0$. 
Note further that since, by assumption, $ - \lambda_{0} + \lambda_{0}^{(0)} = \mathcal O(1/L^2)$,
it leads to a suppressed contribution to $\lambda_0^{(3)}$
of  $\mathcal O(1/L^4)$. This can be dropped. 

We next consider the regime in which both momenta are large, of $ \mathcal O(L^0)$. 
Here $F^{i\epsilon}$ is exponentially suppressed, and can be dropped.
The contribution  involving three factors of $\slashed G$ comes with
three explicit factors of $1/L^3$,
but two of these are canceled by the sums over $\vec k$ and $\vec p$.
Thus, as in the small-momentum regime,
this term is ${\cal O}(1/L^3)$, but in this regime all partial waves must be kept.
This leaves the term containing $ - \lambda_{0} + \lambda_{0}^{(0)} $ and two factors
of $\slashed G$. Since  $- \lambda_{0} + \lambda_{0}^{(0)} = \mathcal O(1/L^2)$, 
this contribution has an explicit factor of $1/L^8$, 
one power larger than the explicit factor on the three-$\slashed G$ term.
However, since $- \lambda_{0} + \lambda_{0}^{(0)}$ is diagonal, 
this contribution is only enhanced only by one sum rather than two, 
leading to an overall $1/L^5$ scaling. Thus this term can also be dropped.

The final region to consider is that in which one momentum is small and the other large. 
Since $F^{i\epsilon}$ and $-\lambda_0+\lambda_0^{(0)}$ are diagonal in momentum space,
this regime is only possible for the term containing three $\slashed G$s.
Since we are keeping this term for all momenta anyway, no special attention to this case is needed. 

Based on these considerations, we deduce that
\begin{equation}
\lambda^{(3)}_{0}  =   \sum_{\vec k  }   \slashed G_{0k}  \tMin{kk} \slashed F^{i \epsilon}_{kk}  \tMi{kk} \slashed G_{k0} +\left [     \slashed G  [- \tM ]   \slashed G [ - \tM  ]    \slashed G   \right ]_{00} + \mathcal O\left(\frac{1}{L^4} \right ) \,.
\label{eq:lam3fin}
\end{equation}
Here the notation in the matrices in the first term indicates that only $\ell = 0$ components
are kept, e.g. $\slashed F_{kk}^{i\epsilon}\equiv\slashed F^{i\epsilon}_{k00;k00}$.
By contrast, the intermediate indices are summed over all momenta
and all partial waves in the second term.
We stress again that the first term is dominated by small momenta, while in the second all
momenta contribute.

The generalization to higher orders is now clear.
For $n>3$ one has four or more factors drawn from $\slashed G$, $\slashed F^{i \epsilon}$ and 
$- \lambda_{0} + \lambda_{0}^{(0)} $. 
This means that the low-momentum contribution scales as $1/L^4$ or higher and can be dropped.
 In the high-momentum regime an ${\cal O}(1/L^3)$ contribution does arise, given by
\begin{equation}
\lambda^{(n)}_{0}   = \left [ \slashed G \left [ - \tM \slashed G \right ]^{n-1} \right ]  
+ \mathcal O\left (\frac{1}{L^4} \right )\,, \ \ \ \ \mathrm{for}\ n >3 \,.
\end{equation}
The $n-1$ momentum sums cancel all but one of the factors of $1/L^3$ contained in
the $\slashed G$s, so that the overall scaling is $1/L^3$.
All other contributions are suppressed.

Summing our results for $\lambda_0$ to all orders, we conclude that
\begin{equation}
\label{eq:finallam0}
\lambda_0= \lambda_0^{(0)}
+  \sum_{\vec k} \slashed G_{0k}  \tMi{kk} \slashed F^{i \epsilon}_{kk}    \tMi{kk}  \slashed G_{k0}  
+ \sum_{n=1}^\infty  \left[ \slashed G  \Big [ - \tM \slashed G \Big ]^n  \right ]_{00}
+ \mathcal O \left (\frac{1}{L^4} \right)
\,,
\end{equation}
where, in the last term, all intermediate momenta and partial waves must be kept.

\bigskip

We turn now to the perturbative analysis of the state $\vert \lambda_0 \rangle$,
using Eq.~(\ref{eq:nthorderstate}).
We are specifically  interested in the two quantities involving this state that enter into
the quantization condition Eq.~(\ref{eq:qcondnomat}).
These are the normalization ${\cal N}_0$ and the matrix element ${\cal Z}$ [Eq.~(\ref{eq:Zdef})].
For both of these, we need only the leading $L^0$ behavior when
$L\to\infty$ with $\Delta E$ tuned such that $\lambda_0\sim 1/L^3$.

The task of identifying the leading terms is similar to that for $\lambda_0$.
After making the simplifications that follow from the
properties of $\slashed G$, $\slashed F^{i\epsilon}$, and $\Rzero$, 
the first two terms can be written
\begin{align}
\vert \lambda^{(1)}_0 \rangle &= 
  [-\tM] (\slashed G + \slashed F^{i\epsilon}) 
\vert \lambda^{(0)}_0 \rangle
  \left[1 + {\cal O}(1/L^2)\right]\,,
\label{eq:firstorderstate}
\\
\vert \lambda^{(2)}_0 \rangle &= 
 [- \tM] (\slashed G + \slashed F^{i\epsilon}-\lambda_0+\lambda_0^{(0)}) 
[- \tM] (\slashed G + \slashed F^{i\epsilon}) 
\vert \lambda^{(0)}_0 \rangle
  \left[1 + {\cal O}(1/L^2)\right] \,.
\label{eq:secondorderstate}
\end{align}
Using these results, we find that the leading order correction to 
${\cal N}_0 = \langle \lambda_0 \vert \lambda_0\rangle$
occurs at second order:\footnote{%
The first-order term vanishes because $\slashed G_{00}=\slashed F^{i\epsilon}_{00}=0$.}
\begin{multline}
{\cal N}_0= 1 +
\langle \vec 0,0,0 \vert 
(\slashed G + \slashed F^{i\epsilon})^\dagger
\tM^\dagger \tM
(\slashed G + \slashed F^{i\epsilon})
\vert \vec 0,0,0\rangle
\\
+
2 {\rm Re}
\langle \vec 0,0,0 \vert \tM
(\slashed G + \slashed F^{i\epsilon})
\tM
(\slashed G + \slashed F^{i\epsilon})
\vert \vec 0,0,0\rangle
+ {\cal O}(1/L^3) \,.
\label{eq:secondorderN0}
\end{multline}
Here we are already using the result that higher-order contributions are of
${\cal O}(1/L^3)$, as will become clear shortly.
Note also that, at this stage, we have to account for the fact, noted above, 
that $\tM$ and $\slashed F^{i\epsilon}$ are not hermitian.
In the low-momentum regime, both of the second-order terms scale as
$1/L^2$, since the dominant terms in $\slashed G$ and $\slashed F^{i\epsilon}$ 
scale as $1/L$. Similarly, at $n$th order, the low-momentum terms scale as $1/L^n$.
In the high-momentum regime, $F^{i\epsilon}$ can be dropped, and each of the
$\slashed G$ factors has an explicit $1/L^3$. There is, however, only a single
intermediate sum over $\vec k$, so the overall scaling is as $1/L^3$.
The same can be easily seen to hold at all higher orders.
We thus conclude that
\begin{equation}
{\cal N}_0 = 1 + {\cal O}(1/L^2)\,.
\end{equation}

Now we turn to the matrix element ${\cal Z}$, which can be expanded
as a geometric series
\begin{equation}
{\cal Z} = 
    \langle \lambda_0 \vert \tilde F   \tKdf  \tilde F \vert \lambda_0  \rangle 
    -
    \langle \lambda_0 \vert \tilde F   \tKdf \overline  F_{3}  \tKdf  \tilde F 
  \vert \lambda_0  \rangle 
  + \cdots \,.
\label{eq:Zexpand}
\end{equation}
Our aim is to substitute the perturbative expansion of $\vert \lambda_0\rangle$
and determine the $L^0$ part of ${\cal Z}$.
We note immediately that the contribution from the
low-momentum regime  in the results
(\ref{eq:firstorderstate}) and (\ref{eq:secondorderstate})
are suppressed by powers of $1/L$ and can be dropped. 
The same is true at higher orders.
In the high-momentum regime the dominant contribution comes from
terms with multiple $\slashed G$s (since, as in the analysis
for $\lambda_0$,  $\slashed F^{i\epsilon}$ is
exponentially suppressed and the $-\lambda_0+\lambda_0^{(0)}$ term lacks
a momentum sum to cancel the explicit $1/L^2$).
This high-momentum contribution is of ${\cal O}(L^0)$ and must be kept.
To see this scaling, 
consider the first term on the right-hand side of Eq.~(\ref{eq:Zexpand})
and substitute Eq.~(\ref{eq:firstorderstate}) for $\vert \lambda_0\rangle$.
The presence of a factor of $\Kdf$ in the ``middle" of
the matrix element implies that there is one momentum
sum for each factor of $\slashed G$, and this cancels the $1/L^3$ factors in $\slashed G$. 
The same cancellation occurs at all orders in perturbation theory,
and also for the higher-order terms in the geometric series in Eq.~(\ref{eq:Zexpand}).
This implies that, in the evaluation of the leading order contribution to ${\cal Z}$, 
we can make the following substitution for the $n$th order term 
\begin{equation}
\vert \lambda^{(n)}_{0} \rangle \longrightarrow
\left[-\tM \slashed G\right]^n \vert \vec 0, 0, 0\rangle
\,.
\end{equation}
These leading terms can then be summed into
\begin{equation}
\vert \lambda_0 \rangle \longrightarrow
\frac1{1+ \tM \slashed G} \vert \vec 0, 0, 0\rangle
\,.
\end{equation}
Thus we find
\begin{equation}
{\cal Z} =
\langle \vec 0, 0, 0\vert \frac1{1+ \slashed G \tM} 
\tilde F   \tKdf  \frac1{1 + \overline F_{3}  \tKdf }  \tilde F 
\frac1{1+ \tM \slashed G} \vert \vec 0, 0, 0\rangle + {\cal O}(1/L)\,.
\label{eq:Zfinal}
\end{equation}
Here we have used the result that $\tM$ is hermitian at ${\cal O}(L^0)$.

\subsection{Relation to the divergence-free three-to-three scattering amplitude}
 \label{sec:MbartoMdf}

In this subsection we demonstrate the following relation between the matrix
element appearing in our modified quantization condition, Eq.~(\ref{eq:qcondnomat}),
and the infinite-volume divergence-free three-to-three scattering amplitude
at threshold [defined in Eq.~(\ref{eq:Mdf3thdef}) below]:
\begin{align}
\label{eq:recoverMdf}
\left\{9 (\tKin{00})^2  {\cal Z}\right\}
\bigg \vert_{E=3m + \DEth} & =  
\Mdfin{00}  
+ \mathcal O(1/L) \,.
\end{align}
This is a key result as it allows us to connect the output of the finite-volume quantization
condition to an infinite-volume scattering quantity.
We stress that this result only holds when the quantity on the left-hand side is evaluated
at $E= 3m + \DEth$, i.e.~the energy must be held at the solution to the quantization condition as $L\to\infty$.

We first review the definition of $\Mdf$, given in Eq.~(87) of Ref.~\cite{KtoM}.
To do so we introduce the set of integrals
\begin{equation}
\label{eq:Indef}
i I_{n; \ell' m'; \ell m}^{(u,u)}(\vec p, \vec k) \equiv \int \frac{d^3 k_n}{(2 \pi)^3 2 \omega_{k_n}} \cdots \int \frac{d^3 k_1}{(2 \pi)^3 2 \omega_{k_1}}  \, \left [ i \mathcal M_2(\vec p \,) \, i G^\infty \big (\vec p, \vec k_n \big) \, i \mathcal M_2(\vec k_n) \cdots i G^\infty \big (\vec k_1, \vec k \, \big) \, i \mathcal M_2(\vec k) \right ]_{\ell' m'; \ell m} \,,
\end{equation}
where $n$ is a positive integer,
\begin{equation}
 G_{\ell' m'; \ell m}^\infty(\vec p, \vec k) = \left(\frac{k^*}{q_p^*}\right)^{\ell'} 
\frac{4 \pi Y_{\ell',m'}(\hat  k^*) 
H(\vec p\,) H(\vec k\,) Y_{\ell,m}^*(\hat p^*)} 
{2 \omega_{kp} (E - \omega_k - \omega_p - \omega_{kp} + i \epsilon)}
\left(\frac{p^*}{q_k^*}\right)^\ell \,,
\label{eq:Ginfdef}
\end{equation}
and
\begin{equation}
\label{eq:Mofkdef}
\mathcal M_{2;\ell',m';\ell,m}(\vec k) \equiv  \delta_{\ell'\ell} \delta_{m'm} 
\left[ \frac{q_k^*}{16 \pi E_{2,k}^*} \cot \delta_{\ell}(q_k^*)  + \tilde\rho(E_{2,k}^*) \right ]^{-1} \,,
\end{equation}
is the standard two-to-two scattering amplitude for two particles carrying 
energy momentum $(E - \omega_k, - \vec k)$. This differs from the matrix $\M$, 
introduced in Eq.~(\ref{eq:M2def}) above, only in that $\mathcal M_2(\vec k)$
is defined for continuous $\vec k$. 
The products in the square brackets of Eq.~(\ref{eq:Indef}) are understood as matrix products 
over the spherical-harmonic indices. 
We also extend the definition to $n=0$ via
\begin{equation}
i I_{0;\ell' m'; \ell m}^{(u,u)}(\vec p, \vec k) \equiv  \left [ i \mathcal M_2(\vec p \,) \, i G^\infty \big (\vec p, \vec k \big) \,  i \mathcal M_2(\vec k) \right ]_{\ell' m'; \ell m} \,.
\label{eq:I0def}
\end{equation}

\begin{figure}
\begin{center}
\includegraphics[scale=0.38]{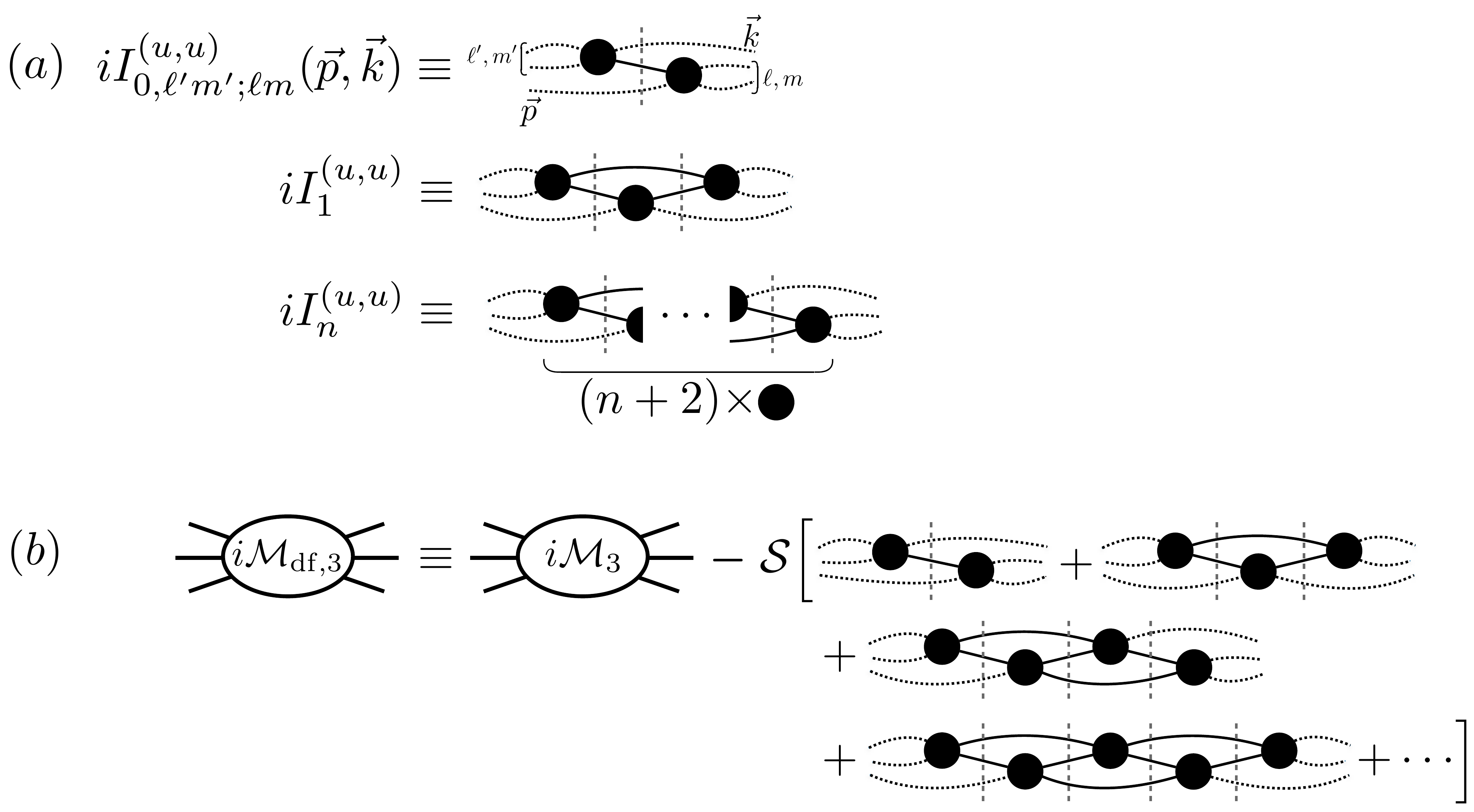}
\end{center}
\caption{Diagrammatic definitions of quantities defined in the text.
(a) The unsymmetrized subtraction functions, $I_{n}^{(u,u)}$. Here the black disks represent on-shell projections of $\mathcal M_2$, and the vertical dashed lines represent simple poles, used in place of the propagators. For $I_0^{}$ we have indicated the coordinate dependence, which applies for all of the functions.  (b) The divergence-free three-to-three amplitude, $\Mdf$. This quantity is given by subtracting an infinite series of pairwise scattering diagrams, $\sum_{n=0}^\infty I_n$, from the standard three-to-three scattering amplitude, $\mathcal M_3$. Here $\mathcal S$ indicates that the symmetrized versions of $I_n$ are to be used in the subtraction.
}
\label{fig:Mdf}
\end{figure}

These definitions are shown diagrammatically in Fig.~\ref{fig:Mdf}(a).
The basic structure is a sequence of on-shell scattering amplitudes
alternating with a pole term that interchanges the scattering pair.
The superscript $(u,u)$ on $ I_{n}^{(u,u)}(\vec k, \vec p)$ indicates that the quantity is
unsymmetrized, in the sense that the momenta $\vec k$ and $\vec p$ are assigned to 
the particles that are unscattered by the outermost insertions. 
The factors of $ G^\infty$ (represented by the vertical dashed lines in the figure)
have the same singularities as propagators in the standard Feynman rules for the diagrams. 
Thus the integrals $ I_{n}^{(u,u)}$ are simplified versions of the corresponding Feynman diagrams,
having the same singularities, but depending only on on-shell two-to-two scattering amplitudes. 

We next define symmetrized versions of these integrals
\begin{equation}
\label{eq:symmdef}
i I_n(\vec p, \hat a'^*; \vec k, \hat a^*) \equiv \mathcal S \big [ i I^{(u,u)}_{n;\ell' m'; \ell m}(\vec p, \vec k) \big ] \equiv \sum_{\{\vec x', \vec y'\} \in \mathcal P_{\mathrm{out}}}  \sum_{\{\vec x, \vec y\} \in \mathcal P_{\mathrm{in}}} 4 \pi Y^*_{\ell'm'}(\hat y'^*) i I_{n;\ell' m'; \ell m}^{(u,u)}(\vec x', \vec x) Y_{\ell m}(\hat y^*) \,,
\end{equation}
where we sum over possible external momentum assignments
\begin{align}
\mathcal P_{\mathrm{out}} &\equiv 
\big \{ \{\vec p, \vec a'\}, \  \{\vec a',  - \vec a' - \vec p\,\}, 
\  \{- \vec a' - \vec p, \vec p \,\} \big \} \,,
\\
\mathcal P_{\mathrm{in}} &\equiv \big \{ \{\vec k, \vec a\}, \  \{\vec a,  - \vec a - \vec k\}, 
\  \{- \vec a - \vec k, \vec k\} \big \}\,.
\end{align}
Here $\vec k$, $\vec a$ and $-\vec a-\vec k$ are the momenta of the initial particles,
while $\vec p$, $\vec a'$ and $-\vec a'-\vec p$ are those of the final particles.
The direction $\hat a^*$ is that of $\vec a$ after boosting to the CM frame
of the scattered pair, with $\hat a'^*$ defined analogously for the final state.
Similarly, when the momentum pair is $\vec x, \vec y$, 
$\hat y^*$  is defined by boosting $(\omega_y, \vec y \, )$ with 
with velocity $\vec \beta = \vec x/(E - \omega_x)$.
Note also that, prior to symmetrization, we have to insert the spherical harmonics
and sum over their indices in order to obtain functions of the external momenta.

As we explain in detail in Refs.~\cite{us,KtoM}, 
the sum over all symmetrized integrals $I_n$ has the same singularities 
as the full three-to-three scattering amplitude $\mathcal M_3$. 
Thus the difference between these quantities, which we denote $\mathcal M_{\mathrm{df},3}$,
is free of divergences. Explicitly, this is defined as
\begin{equation}
\label{eq:Mdfdef}
i \mathcal M_{\mathrm{df},3}(\vec p, \hat a'^*; \vec k, \hat a^*) \equiv  i \mathcal M_{3}(\vec p, \hat a'^*; \vec k, \hat a^*) - \sum_{n=0}^\infty iI_n(\vec p, \hat a'^*; \vec k, \hat a^*) \,,
\end{equation}
and is shown diagrammatically in Fig.~\ref{fig:Mdf}(b). 
What is required for Eq.~(\ref{eq:recoverMdf}) is the value of this amplitude at threshold:
\begin{equation}
\Mdfin{00}
 \equiv \mathcal M_{\rm df,3}(\vec 0,\hat a'^*;\vec 0,\hat a^*)\big\vert_{E=3m}
\,.
\label{eq:Mdf3thdef}
\end{equation}
Note that the right-hand side is, in fact, independent of
the direction vectors $\hat a'^*$ and $\hat a^*$, since $\vec a'^*=\vec a^*=0$ 
when $\vec p=\vec k=0$ and $E=3m$.
Thus we have included no such dependence in $\Mdfin{00}$. 
An equivalent definition is giving by decomposing 
$\Mdf(\vec p, \hat a'^*; \vec k, \hat a^*)$ in spherical harmonics, 
keeping only the $s$-wave term, and evaluating this at threshold. 
Thus the index label on $\Mdfin{00}$ is consistent with that for 
$\tK$, $\tilde G$ and $\tilde F$ used above.

Having explained the definition of the right-hand side of Eq.~(\ref{eq:recoverMdf})
we now turn to proving the claim. To do so we write out the two sides in more detail.
Using the result for ${\cal Z}$ worked out in the previous subsection, Eq.~(\ref{eq:Zfinal}),
we can express the left-hand side of (\ref{eq:recoverMdf}) as
\begin{align}
\left\{9 (\tKin{00})^2  {\cal Z}\right\} \bigg \vert_{E=3m + \Delta E} 
& =  
 \overline L  \; \tKdf \frac{1}{1\!+\!  \overline  F_{3}  \tKdf}   \; \overline R \bigg \vert_{E=3m + \Delta E}
 + {\cal O}(1/L)
 \,.
 \label{eq:newlhs}
 \end{align}
Here we have introduced the row and column vectors
\begin{align}
 \overline L_{k\ell m} = - 3  \tKin{00} \left[ \frac1{1+\slashed G \tM} \tilde F \right]_{000,klm}\,,
\label{eq:Lbardef}
\\
 \overline R_{k\ell m} = - 3 \left[\tilde F \frac1{1+\tM \slashed G}\right]_{klm,000} \tKin{00}  \,.
\label{eq:Rbardef}
\end{align}

As for the right-hand side, we can rewrite this using the general relation
between finite-volume and infinite-volume three-particle scattering
amplitudes given in Eq.~(80) of Ref.~\cite{KtoM}. 
For the threshold amplitude this relation is
\begin{equation}
\label{eq:originalKtoMrel}
\Mdfin{00} = \lim_{L \rightarrow \infty} \bigg \vert_{i \epsilon}  
 \mathcal S\bigg [ 
 \left(\frac13 - \frac1{1+\mathcal M_{2,L} G} \mathcal M_{2,L} F\right)
 \Kdf \frac{1}{1\!+\!  F_3  \Kdf} 
  \left(\frac13 - \frac{F}{2\omega} \frac1{1+G\mathcal M_{2,L}} \mathcal M_{2,L} (2\omega)\right)
\bigg ]_{000;000} \,,
\end{equation}
where $E=3m$, and the new matrix is $\mathcal M_{2,L}^{-1}\equiv \K^{-1} + F$.
Note that here we are apparently taking a step backwards by expressing the
infinite-volume quantity $\Mdfin{00}$ in terms of the $L\to\infty$ limit of
a finite-volume matrix element. The reason for doing so is that the
connection to the left-hand side of the desired relation (\ref{eq:recoverMdf})
is then much clearer.
One new feature in (\ref{eq:originalKtoMrel}) is that the infinite-volume limit
is taken using an $i\epsilon$ prescription. As explained in Ref.~\cite{KtoM},
a prescription is needed to avoid singularities in $F$ and $G$.
The prescription that is required for (\ref{eq:originalKtoMrel}) to hold is that 
singularities in summands are shifted by $i\epsilon$, after which the infinite-volume
limit is well defined.

The symmetrization operator, ${\cal S}$, in Eq.~(\ref{eq:originalKtoMrel}) is
essentially the same as that defined in Eq.~(\ref{eq:symmdef}),
although there are some subtleties when applied to the finite-volume
matrices~\cite{KtoM}.
These do not concern us here, however, because symmetrization is trivial for
the threshold amplitude---it leads simply to an overall factor of 9.

We proceed by rewriting the result (\ref{eq:originalKtoMrel}) in a form that
is similar to Eq.~(\ref{eq:newlhs}). After some reorganization (including insertions
of appropriate factors of $Q Q^{-1}$ and using $Q_{00}=1$) we find
\begin{equation}
\label{eq:generalKtoMrel}
\Mdfin{00}
 = \lim_{L \rightarrow \infty} \bigg \vert_{i \epsilon}  
\tilde L \;
 \tKdf \frac{1}{1\!+\! \tilde F_3  \tKdf} 
\;\tilde R   \,,
\end{equation}
where $E=3m$ and the new row and column vectors are
\begin{align}
\tilde L_{k\ell m} &= 
   \bigg [ 1- 3 \mathcal H^{-1}   \; \tilde  F \bigg ] _{000,k \ell m} 
 \label{eq:Ltildedef} \,, \\
\tilde R_{k \ell m} & =  
\bigg [ 1 - 3 \tilde  F \mathcal H^{-1}  \;  \bigg ] _{k \ell m,000} 
 \,.
\label{eq:Rtildedef}
\end{align}
The result we are aiming to demonstrate can now be rewritten as
\begin{equation}
\lim_{L \rightarrow \infty}
\overline L \;  \tKdf \frac{1}{1\!+\!  \overline  F_{3}  \tKdf}   \; \overline R \bigg \vert_{E=3m + \DEth}
 = \lim_{L \rightarrow \infty} \bigg \vert_{i \epsilon}  
\tilde L \;
 \tKdf \frac{1}{1\!+\! \tilde F_3  \tKdf} 
\;\tilde R\bigg\vert_{E=3m}   \,.
\end{equation}
We have chosen the notation in such a way that the results look similar, but we still have
significant work to do to demonstrate equality.
We stress that the nature of the infinite-volume limits differs between the two sides:
on the left-hand side $\Delta E$ is tuned to satisfy the quantization condition,
while on the right-hand side $\Delta E=0$.

We first focus on the matrices between the ``L'' and ``R'' vectors, and show that 
\begin{equation}
\label{eq:FKFall}
\lim_{L \rightarrow \infty}  \tKdf\frac1{1+\overline F_3 \tKdf} \bigg \vert_{E = 3m + \DEth}  =
\lim_{L \rightarrow \infty} \bigg \vert_{i \epsilon} 
\tKdf \frac1{1+ \tilde F_3 \tKdf }\bigg \vert_{E = 3m} \,.
\end{equation}
We first give a qualitative explanation of this equality.
The limit on the right-hand side
has been investigated in Ref.~\cite{KtoM}, and is given by an
infinite-volume function ${\cal T}_{\ell',m';\ell, m}(\vec p, \vec k)$
[where $\vec p$ and $\vec k$ are the external spectator momenta,
both held fixed in the limit]. The contribution that survives in the 
limit comes from large intermediate momenta---contributions
from low momenta are suppressed by powers of $1/L$.
We note also that, once the limit is taken, we can send $\epsilon\to 0^+$, since the
poles are at threshold, 
and do not need regulation once sums have been converted to integrals.
The quantity on the left-hand side differs in two ways:
(i) $\tilde F_3$ is replaced by $\overline F_3=\tilde F_3- F_3^{\lambda_0}$, and
(ii) the infinite-volume limit is approached with the tuned $\DEth = {\cal O}(1/L^3)$
rather than $\Delta E=0$. Note that this limit avoids the poles
in $F$ and $G$ that occur at $\Delta E=0$, so that one does not need to use the
$i\epsilon$ prescription at an intermediate stage. 
Thus, as far as the contributions from large intermediate momenta are concerned,
one approaches exactly the same kinematic point as on the right-hand side and
should attain the same limit. For large momenta the subtracted part $F_3^{\lambda_0}$
is suppressed by powers of $1/L$.
The only complication is that, when approaching the limit with tuned $\Delta E$,
there is an enhanced low-momentum contribution to $\tilde F_3$, namely 
that from $F_3^{\lambda_0}$. This, however, is removed by the subtraction in $\overline F_3$, 
so the left-hand side also receives no low-momentum contributions as $L \to \infty$.

To demonstrate the result in detail we expand both sides of (\ref{eq:FKFall})
in a geometric series and argue that the results agree order by order.
The leading order terms are identical,
so the first nontrivial result to show is
\begin{equation}
\label{eq:FKFlimitident}
\lim_{L \rightarrow \infty}  \tKdf (\tilde F_3-F_3^{\lambda_0}) \tKdf \bigg \vert_{E = 3m + \DEth} 
=
\lim_{L \rightarrow \infty} \bigg \vert_{i \epsilon} \tKdf \tilde F_3 \tKdf \bigg \vert_{E = 3m} \,.
\end{equation}
Our first step is to replace the $i\epsilon$ regulated limit on the right-hand side
with one in which $E-3m=c/L^3$, with $c$ any constant differing from the tuned value
$a_3$ to be determined below. This avoids the poles in $F$ and $G$
[which are at $E=3m$ and $E=3m + {\cal O}(1/L^2)$] so that the limit is well-defined.
In other words we have
\begin{equation}
\lim_{L \rightarrow \infty} \bigg \vert_{i \epsilon} \tKdf \tilde F_3 \tKdf \bigg \vert_{E = 3m} 
= 
\lim_{L \rightarrow \infty}  \tKdf \tilde F_3 \tKdf \bigg \vert_{E = 3m + c/L^3} \,.
\label{eq:step1}
\end{equation}

Next we argue that on the left-hand side of (\ref{eq:FKFlimitident}) we can
replace $E=3 m+\DEth$ with $E=3m+c/L^3$, with any choice of $c$:
\begin{equation}
\lim_{L \rightarrow \infty}  \tKdf (\tilde F_3-F_3^{\lambda_0}) \tKdf \bigg \vert_{E = 3m + \DEth} 
=
\lim_{L \rightarrow \infty}  \tKdf (\tilde F_3-F_3^{\lambda_0}) \tKdf \bigg \vert_{E = 3m + c/L^3} 
\,.
\label{eq:step2}
\end{equation}
Note here that 
$F_3^{\lambda_0}=-\tilde F|\lambda_0\rangle\langle\lambda_0|\tilde F/({\cal N}_0 L^3\lambda_0)$
depends on how $E$ is chosen to approach $3m$, 
since both $\lambda_0$ and the $|\lambda_0\rangle$ depend on $E$.
To understand this equality consider first the $\tilde F_3$ terms.
We recall from our earlier discussion that $\tilde F_3$ has an explicit factor of
$1/L^3$, whereas $\tKdf\sim {\cal O}(L^0)$. The $1/L^3$ can only be canceled
by a sum over large intermediate momenta 
(leading to the infinite volume function ${\cal T}$ described above)
or by the presence of an eigenvalue of ${\cal H}$ scaling as $1/L^3$. 
The latter corresponds to a low-momentum intermediate state since
$|\lambda_0\rangle = |\vec 0,0,0\rangle + {\cal O}(1/L)$.
The subtraction on the left-hand side removes this potential ${\cal O}(L^0)$
contribution, however, so that the difference $\overline F_3$ cannot
give rise to an ${\cal O}(L^0)$ low-momentum contribution.
Thus it makes no difference precisely how the large volume limit
is taken as long as the same asymptote is approached.
This is the case for the two sides of (\ref{eq:step2}) for any choice of $c$.

Finally, we note that the $F_3^{\lambda_0}$ term can be dropped from the right-hand side
of Eq.~(\ref{eq:step2}),
\begin{equation}
\lim_{L \rightarrow \infty}  \tKdf (\tilde F_3-F_3^{\lambda_0}) \tKdf \bigg \vert_{E = 3m + c/L^3} 
=
\lim_{L \rightarrow \infty}  \tKdf \tilde F_3 \tKdf \bigg \vert_{E = 3m + c/L^3} 
\,,
\label{eq:step3}
\end{equation}
as long as $c\ne a_3$.
This is simply because the explicit factor of $1/L^3$ in $F_3^{\lambda_0}$ cannot be canceled
for an untuned energy.

Combining these three steps we find that the left- and right-hand sides of
(\ref{eq:FKFlimitident}) are equal.
This argument can be extended almost verbatim to the higher order terms in
the expansions of Eq.~(\ref{eq:FKFall}), and we do not repeat the discussion.
This establishes the desired equality, Eq.~(\ref{eq:FKFall}).

It remains only to relate the ``end caps'' that appear in Eqs.~(\ref{eq:newlhs})
and (\ref{eq:generalKtoMrel}).
We consider first the barred end caps of Eqs.~(\ref{eq:Lbardef}) and (\ref{eq:Rbardef}),
which are to be evaluated along the tuned energy trajectory $E=3m+\DEth$
in the limit $L\to\infty$.
This means that we can replace $\tKin{00}$ with $\tMin{00}$, and that,
as noted above following Eq.~(\ref{eq:xdef}), the combination
 $-3\tMin{00}\tilde F_{00}$ has the limiting value of unity.
However, $\slashed G$ does not  contain an ${\cal O}(L^0)$ term when $\Delta E\to 0$, 
since the potentially large term has been subtracted.
Combining these observations we find
\begin{align}
\lim_{L\to\infty} \overline L_{k\ell m}\bigg\vert_{E=3m+\DEth} &= 
 \left [ - 3  \tM \tilde F + 3  \tM \slashed G  \tM  \frac{1}{1+\slashed G \tM}  \tilde F \right ]_{000,k \ell m} 
=
\left [1 + 3  \tM \slashed G  \tM  \frac{1}{1+\slashed G \tM}    \tilde F \right ]_{000,k \ell m} 
   \,, \\
\lim_{L\to\infty}  \overline R_{k\ell m}\bigg\vert_{E=3m+\DEth} & = 
  \left [ - 3  \tilde F  \tM + 3 \tilde F \tM \slashed G  \tM  \frac{1}{1+\slashed G \tM}  \right ]_{k \ell m,000}  
=  \left [ 1 + 3 \tilde F \tM \slashed G  \tM  \frac{1}{1+\slashed G \tM}     \right ]_{k \ell m,000}  
\,,
\end{align}
where we have left the infinite-volume limit and the constraint $E=3m+\DEth$ implicit in the middle and final equality.

Turning to the ``tilded" end caps of Eqs.~(\ref{eq:Ltildedef}) and (\ref{eq:Rtildedef}), the
infinite-volume limit is to be taken with $E=3m$ using the $i\epsilon$ prescription. 
This means that the enhanced eigenvalue of
${\cal H}= \tM^{-1}+ \tilde F^{i\epsilon} + \tilde G$ plays no role.
As explained in Ref.~\cite{KtoM}, $\tilde F^{i\epsilon}$ vanishes in this limit
[since it is a difference between a sum and an integral regulated using
an $i\epsilon$ prescription, see Eq.~(\ref{eq:Fdef1})].
However, $\tilde F$ does not vanish in general, due to the contribution of the $\rho$
term [see Eq.~(\ref{eq:Fdef2})], although $\tilde F_{00}$ does vanish at threshold,
since $\rho$ vanishes there. We find
\begin{align}
\lim_{L \rightarrow \infty} \bigg \vert_{i \epsilon}  \tilde L_{k\ell m} &= 
 \lim_{L \rightarrow \infty} \bigg \vert_{i \epsilon}   \bigg [ 1- 3  \tM  \frac{1}{1 + \tilde  G \tM}     \tilde  F \bigg ] _{000,k \ell m}  =\lim_{L \rightarrow \infty} \bigg \vert_{i \epsilon}    \bigg [ 1+ 3  \tM \tilde  G \tM \frac{1}{1 + \tilde  G \tM}     \tilde  F \bigg ] _{000,k \ell m} \,, \\
\lim_{L \rightarrow \infty} \bigg \vert_{i \epsilon}  \tilde R_{k \ell m} & =  \lim_{L \rightarrow \infty} \bigg \vert_{i \epsilon} 
\bigg [ 1 - 3 \tilde  F   \tM  \frac{1}{1 + \tilde  G \tM}    \bigg ] _{k \ell m,000} = \lim_{L \rightarrow \infty} \bigg \vert_{i \epsilon} \bigg [ 1 + 3 \tilde  F   \tM \tilde  G \tM \frac{1}{1 + \tilde  G \tM}    \bigg ] _{k \ell m,000} \,.
\end{align}
To complete the argument we note that the distinction between $\tilde G$ and $\slashed G$ is subleading in $L$.
We deduce
\begin{equation}
\lim_{L \rightarrow \infty}
\overline L \, \Big \vert_{E=3m + \DEth}
 = \lim_{L \rightarrow \infty} \bigg \vert_{i \epsilon}  
\tilde L \,  \Big \vert_{E=3m}   \,, \ \ \ \ \ \ \ \ \ \   \lim_{L \rightarrow \infty}
\overline R \, \Big \vert_{E=3m + \DEth}
 = \lim_{L \rightarrow \infty} \bigg \vert_{i \epsilon}  
\tilde R \,  \Big\vert_{E=3m}   \,.
\label{eq:equalends}
\end{equation}
Combining Eqs.~(\ref{eq:FKFall}) and (\ref{eq:equalends}) completes the 
demonstration of the desired result, Eq.~(\ref{eq:recoverMdf}).

\subsection{Relation to minimally subtracted threshold three-to-three amplitude}
\label{sec:MdftoMthr}

Using the results (\ref{eq:finallam0}) and (\ref{eq:recoverMdf}),
as well as the equality of $\tKin{00}$ and $\tMin{00}$ at threshold,
we can rewrite the quantization condition (\ref{eq:qcondnomat}) as
\begin{multline}
 9 L^3 \Bigg\{
(\tKin{00})^2 \lambda_0^{(0)}
+\left[ [-\tM] \slashed G \sum_{n=1}^\infty    \Big [ - \tM \slashed G \Big ]^n  [-\tM] \right ]_{00} 
\\
+  \sum_{\vec k} \tMi{00}\slashed G_{0k}  \tMi{kk} \slashed F^{i \epsilon}_{kk}    \tMi{kk}  \slashed G_{k0} \tMi{00}
\Bigg\} \Bigg\vert_{E=3m+\DEth} 
= \Mdfin{00} + \mathcal O(1/L)
\,.
\label{eq:qcondnew}
\end{multline}
We recall that the second term in curly braces 
contains low-momentum contributions scaling as $1/L^2$ and $1/L^3$,
and a high-momentum contribution scaling as $1/L^3$,
while the third term contains only a low-momentum contribution scaling as $1/L^3$.
At this stage we could pull out these low-momentum contributions, evaluate them
explicitly, and replace the high-momentum contributions by appropriate 
infinite-volume integrals. With these expressions in hand we could then
determine the coefficients in the expansion (\ref{eq:DeltaEexp}) for $\DEth$.
The coefficient $a_6$ would then depend on the divergence-free
amplitude at threshold, $\Mdfin{00}$. 

However, there is one feature of such a result that is unsatisfactory.
We recall that $\Mdf$ is defined by subtracting from $\mathcal M_3$
a series of integrals ${ I}_n$ that remove the physical divergences
[see Fig.~\ref{fig:Mdf} and Eq.~(\ref{eq:Mdfdef})].
The issue is that these integrals, defined in Eq.~(\ref{eq:Indef}), 
involve the two-particle scattering amplitude $\M$ evaluated far below threshold
(since the spectator momenta range up to $k\sim m$ at which point the
CM energy of the nonspectator pair is $(3m-\omega_k)^2- k^2 \ll 4 m^2$).
While there is nothing wrong in principle with this (one can obtain the subthreshold
amplitude by analytic continuation) 
it introduces what seems to be an unnecessary complication.
The point of the subtractions, after all, is to remove the physical divergences, which
occur at threshold.

It turns out, however, that the formalism, and in particular, Eq.~(\ref{eq:qcondnew}),
is hinting at a remedy. The high-momentum part of the second term in curly braces
turns out, as shown below, to exactly cancel the high-momentum (far subtheshold)
parts of the integrals ${ I}_n$ contained in $\Mdfin{00}$.
Thus we are led to a different definition of the subtracted threshold amplitude
that depends only on physical quantities at or above threshold.
This is the threshold amplitude $\Mthr$ defined schematically
in the Introduction. Here we give its precise definition and then use it to 
simplify the quantization condition.

\bigskip
Our specific definition of $\Mthr$ makes use of the observation 
that the infinite series of terms subtracted in Eq.~(\ref{eq:Mdfdef}) is not needed 
to reach a divergence-free quantity when working with degenerate particles.\footnote{%
The set of integrals that needs to be subtracted is larger if the particles are not degenerate.
See Refs.~\cite{Rubin:1966zz,us} for more discussion.}
From the general considerations of Ref.~\cite{Rubin:1966zz} 
we know that, above threshold, only $I_0$ and $I_1$ need to be subtracted.
Infrared (IR) divergences are more severe at threshold,
but, as shown in Appendix~\ref{app:Inarefin}, 
$I_n$ with $n\geq3$ remain finite, so the only additional subtraction we need 
at threshold is of $I_2$.
In total, then, our first step towards a definition of $\Mthr$ is to drop the
subtraction of $I_{n}$ with $n\geq3$ from $\Mdfin{00}$. 
The next step is to modify the remaining quantities, 
$I_0$, $I_1$ and $I_2$, to remove the dependence on subthrshold $\M$. 
In fact, since $I_0$ does not contain an integral [see Eq.~(\ref{eq:I0def})], 
we need only to modify the latter two. 

These considerations lead to the definition
\begin{equation}
\label{eq:Mthrdef}
 \Mthr \equiv \lim_{\delta \rightarrow 0} \left[  
 \mathcal M_{3,\delta}(0,\hat a'^*; 0, \hat a^*) - 
 I_{0;\delta}(0,\hat a'^*; 0, \hat a^*) -   
 \int_{\delta}   \frac{d^3 k_1}{(2 \pi)^3}  \Xi_1(\vec k_1) -   
 \int_{\delta}   \frac{d^3 k_1}{(2 \pi)^3}    \int_{\delta}   \frac{d^3 k_2}{(2 \pi)^3}  
 \Xi_2(\vec k_1, \vec k_2)   \right ] \,.
\end{equation}
Here $\delta$ indicates the presence of an IR regularization, to be defined shortly,
while $\Xi_1$ and $\Xi_2$ are the modified integrands whose integrals
replace $I_1$ and $I_2$, respectively.
They are given in Eqs.~(\ref{eq:Xi1def}) and (\ref{eq:Xi2def}) below,
and  depend only on the scattering length, $a$, i.e.~not on the
scattering amplitude for subthreshold momenta.

We begin our explanation of the definition of $\Mthr$ by
describing the $\delta$ regularization. This consists of two parts.
The first is that all IR divergent integrals are cutoff by a lower limit,
$k \geq \delta$ (applied in the frame in which $\vec P=0$).
This is indicated by the subscript on the integrals in Eq.~(\ref{eq:Mthrdef}).
As discussed below, this allows us to set $E=3m$ for these terms, i.e.~to
work directly at threshold.
However, $I_0$ diverges at threshold when
the spectator momenta $\vec p$ and $\vec k$ vanish: 
\begin{equation}
I_0^{(u,u)}(\vec 0, \vec 0) \propto \frac1{E-3m}\,.
\end{equation}
Thus we must introduce a second part in the definition of  $\delta$ regularization:
the energy $E$ must approach threshold as $E-3m \propto \delta^4$
with a nonzero proportionality constant.
The subscript $\delta$ on $I_0$ in Eq.~(\ref{eq:Mthrdef}) indicates that $E-3m$ scales
in this way.
As we explain shortly,
the choice of the fourth power of $\delta$ allows us to effectively work at threshold for
$I_1$ and $I_2$ while regulating $I_0$. 
In fact, any power of $\delta$ greater than cubic suffices.\footnote{%
 Note that, whatever power one chooses, the square of the scattering
 particle momentum in $\mathcal M_2$ within $I_0$ will scale in the
 same way as the energy difference, $q^2 \sim E-3m$. Thus, in the
 $\delta \to 0$ limit, both the scattering length and the effective
 range contribute to the $I_0$ subtraction. Indeed, since the $r$-dependent 
 terms are finite, one could choose not to subtract
 these. This would change the definition of $\Mthr$ and would also
 change the explicit $r$-dependent term in $a_6$ [see Eq.~(\ref{eq:a6res})] to compensate.}

We next determine the form of the modified integrand $\Xi_1$. We begin with
the unsymmetrized form of $I_1$, which is
\begin{equation}
\label{eq:I1def}
i I_{1;\ell',m';\ell,m}^{(u,u)}(\vec p; \vec k)\equiv  
\int \frac{d^3 k_1}{(2 \pi)^3 2 \omega_{k_1}}  \, 
\left[i \mathcal M_2(\vec p \,) \, i G^\infty \big (\vec p, \vec k_1 \big) \, i \mathcal M_2(\vec k_1)
 i G^\infty \big (\vec k_1, \vec k \, \big) \, i \mathcal M_2(\vec k)\right]_{\ell',m';\ell,m} \,.
\end{equation}
We want to pull out from this integral the part that leads to the IR divergence at threshold, 
for this is the only part that we need to subtract from $\mathcal M_3$.
As explained in Appendix~\ref{app:Inarefin}, IR divergences at threshold are present
only if $\vec p=\vec k=0$ and if all three of the scattering amplitudes $\mathcal M_2$
are in the $s$-wave.
Thus we focus on 
\begin{align}
\label{eq:I1IRdefA}
I_{1;00;00}^{(u,u)}(\vec 0; \vec 0)&=
\int \frac{d^3 k_1}{(2 \pi)^3} 
  \frac{ \mathcal M_{2,s}(\vec k_1)}{2 \omega_{k_1}}  \, 
\left( 
 \frac{\mathcal M_{2,s}(\vec 0 \,) H(\vec k_1)}{2\omega_{k_1}(E-m-2\omega_{k_1} + i\epsilon)}\right)^2
+ \textrm{IR finite}\,,
\end{align}
where we are using the abbreviation $\mathcal M_{2,s} \equiv \mathcal M_{2;00;00}$.
The ``IR finite'' term is IR finite at threshold and comes from higher intermediate partial waves.
The integral in (\ref{eq:I1IRdefA}) has a double pole at $k_1=|\vec k_1|= q$, where $q$
[defined in Eq.~(\ref{eq:E20q0})] is the three-momentum of each particle in the nonspectator pair.
This pole is regulated by the $i\epsilon$ prescription that comes with $G^\infty$.
However, unlike the case of a single pole, 
the integral here diverges when $\epsilon\to 0$, for any $E\geq 3m$. 
This divergence is necessary to cancel the corresponding physical divergence in $\Mth$.
The issue at hand is to find a simpler integral that has the same IR divergence at threshold but does
not depend, as $I_1$ does, on $\mathcal M_{2,s}(\vec k_1)$ evaluated far below threshold.

To do so we apply the $\delta$ regularization to $I_1$. Then, in the IR regime where
$k_1\sim \delta$, we have that 
\begin{equation}
E-m-2\omega_{k_1}= -\frac{k_1^2}{m} + E-3m + {\cal O}(k_1^4)
= -\frac{k_1^2}{m}\left[1 + {\cal O}(\delta^2) \right]\,,
\end{equation}
since $E-3m$ scales in the same way as the $k_1^4$ term.
This implies that the pole always lies below the cutoff on $k_1$, so that the
integral is well regulated.
Since the overall IR divergence is linear ($\int dk_1/k_1^2$) the ${\cal O}(\delta^2)$ terms
lead to IR-finite corrections and thus can be dropped from the subtraction to $\mathcal M_3$.
This is why our $\delta^4$ scaling of $E-3m$ is effectively the same as setting $E=3m$.
The same holds for $I_2$, since this integral has a weaker IR divergence.

We conclude that to obtain the same IR divergence as in $I_1^{(u,u)}$ we need only expand
the residue of the double pole about $\vec k_1=0$ and keep the constant
and linear terms. Since $E-3m$ scales quartically we can set $E=3m$ in this expansion.
Similarly we can set $\omega_{k_1}$ to $m$.
The factors of $H(\vec k_1)$ equal unity to all orders in a Taylor expansion about threshold,
but we do not expand them as they are needed for UV convergence in some terms.
Thus all we need to expand is $\mathcal M_{2,s}(\vec k_1)$,
which can be done using the relation between $\tM$ and $\tK$ [Eq.~(\ref{eq:M2def})], the
definition of $\rho$ [Eq.~(\ref{eq:rhodef})], the near-threshold form of
$\K$ [Eqs.~(\ref{eq:Ktwodef}) and (\ref{eq:psexp})], and the expression for
$q_k^2$ [Eq.~(\ref{eq:qstarexp})]. The net result is that the modified integrand is\footnote{%
This factor of $H^2$ is not necessary to make the $1/k_1^4$ term UV convergent,
but we keep it for the sake of uniformity, since the UV cutoff is needed for the $1/k_1^3$ term.}
\begin{equation}
\label{eq:X1uudef}
\Xi^{(u,u) }_1(\vec k_1)  \equiv -  \frac{[32\pi m a]^3}{8m} \left [  
 \frac{H(\vec k_1)^2}{k_1^4}
 +   a\frac{\sqrt{3}}{2}  \frac{H(\vec k_1)^3}{k_1^3} \right ] \,,
\end{equation}
and this satisfies
\begin{equation}
\lim_{\delta \rightarrow 0} \left [ I_{1;\delta;00;00}^{(u,u)}(\vec 0, \vec 0)
-   \int_{\delta}   \frac{d^3 k_1}{(2 \pi)^3}  \Xi_1^{(u,u)}(\vec k_1)  \right ] 
 = \mathrm{finite}\,.
\end{equation}
At threshold, symmetrization leads only to multiplication by 9, so we can replace
the subtraction of $I_1$ with that of the integral of 
\begin{equation}
\Xi_1(\vec k_1)=9\;\Xi_1^{(u,u)}(\vec k_1) \,.
\label{eq:Xi1def}
\end{equation}
This is the quantity entering Eq.~(\ref{eq:Mthrdef}).

A similar analysis for $I_2$ leads to the modified integrand
\begin{align}
\Xi_2(\vec k_1, \vec k_2) & =
\frac{9}{16m^2}  [32 m \pi a]^4  \frac{H(\vec k_1)^2H(\vec k_2)^2}
{k_1^{\,2} [ k_1^{\,2}+k_2^{\,2}+(\vec k_1+\vec k_2)^2] k_2^{\,2} } \,.
\label{eq:Xi2def}
\end{align}
There is only a single term since the integral is only logarithmically IR divergent.

This completes the explanation of the quantities entering the definition of $\Mthr$, Eq.~(\ref{eq:Mthrdef}).
To use this to simplify the quantization condition, we need to relate $\Mthr$ to our original threshold amplitude,
$\Mdfin{00}$. Combining the definition of $\Mdfin{00}$, given
in Eqs.~(\ref{eq:Mdfdef}) and (\ref{eq:Mdf3thdef}), with the result (\ref{eq:Mthrdef}) we find
\begin{equation}
\Mthr = \Mdfin{00} + \lim_{\delta \rightarrow 0} 
\left\{ 
\left(I_{1;\delta}  -   \int_{\delta}   \frac{d^3 k_1}{(2 \pi)^3}  \Xi_1(\vec k_1)\right)
+ \left( I_{2;\delta} -  
\int_{\delta}   \frac{d^3 k_1}{(2 \pi)^3}    \int_{\delta}   \frac{d^3 k_2}{(2 \pi)^3}  \Xi_2(\vec k_1, \vec k_2) \right)
\right \} 
+  \sum_{n=3}^\infty I_{n}  
\,.
\label{eq:M3thtoMdf3th}
\end{equation}
Since $I_0$ does not appear, we can set $E=3m$ in the expression in curly braces. 
In other words, IR regularization is achieved here simply by cutting off the IR divergent integrals.
We are also adopting the notation that $I_n$ or $I_{n;\delta}$ without arguments implies
that both spectator momenta vanish and $E=3m$, so that these are purely $s$-wave quantities
(as for $\Mdfin{00}$).
The interpretation of the result (\ref{eq:M3thtoMdf3th}) is that the
subtraction of $\sum_{n=3}^\infty I_n$ is unnecessary for degenerate particles, 
and so we undo this by adding the series back in. In addition we add back part of $I_1$ and $I_2$, 
but with a subtraction defined using $\Xi_1$ and $\Xi_2$ that keeps $\Mthr$ finite.
\bigskip

We conclude this section by rewriting the quantization condition (\ref{eq:qcondnew}) in terms
of $\Mthr$. We will need the following results
\begin{gather}
\label{eq:Mthrident1}
- 9  L^3  \left[ \tM  \slashed G  \Big [ - \tM \slashed G \Big ]^n \tM \right ]_{00}\Bigg\vert_{E=3m+\DEth}
= I_{n} + \mathcal O(1/L) \ \ \mathrm{for} \ \ n \geq 3 \,, \\
\begin{split}
\label{eq:Mthrident2}
- 9  L^3   \left[ \tM  \slashed G  [- \tM  \slashed G]^2 \tM  \right ]_{00}\Bigg\vert_{E=3m+\DEth}
 - \frac{1}{L^6} \sum_{\vec k_1, \vec k_2 \neq 0} \Xi_2(\vec k_1, \vec k_2) & =\\
 &  \hspace{-100pt} \lim_{\delta \rightarrow 0} \left [ I_{2;\delta} -   \int_{\delta}   
\frac{d^3 k_1}{(2 \pi)^3}    \int_{\delta}   \frac{d^3 k_2}{(2 \pi)^3}  
\Xi_2(\vec k_1, \vec k_2)  \right ]+ \mathcal O(1/L)\,,
\end{split} \\
\begin{split}
- 9  L^3  \left[  \tM \slashed G  [- \tM   \slashed G]   \tM \right ]_{00}\Bigg\vert_{E=3m+\DEth}
-  \frac{1}{L^3} \sum_{\vec k_1  \neq 0} \Xi_1(\vec k_1) + 9 \widetilde {\mathcal K}^2_{2,00} \frac{2 \pi a}{m^3} \frac{\mathcal K x}{(2 \pi)^6} & = \\
  &  \hspace{-100pt} \lim_{\delta \rightarrow 0} \left [ I_{1;\delta} -   \int_{\delta}   
\frac{d^3 k_1}{(2 \pi)^3}  \Xi_1(\vec k_1)  \right ]+ \mathcal O(1/L) \,,
 \label{eq:Mthrident3}
 \end{split}
\end{gather}
which are demonstrated below. The final term on the left-hand side of Eq.~(\ref{eq:Mthrident3}) depends both on the threshold K matrix, $\widetilde{\mathcal K}_{2,00}$, and the geometric constant $\mathcal K$ first introduced in Eq.~(\ref{eq:tildeF00final}) above and discussed in detail in Appendix \ref{app:tF00}. In this term we have also used the parameter $x=\Delta E_{\text{th}} L^3 m^2$, introduced in Eq.~(\ref{eq:xdef}).
 Using these results, and the relation (\ref{eq:M3thtoMdf3th}), we find
that the quantization condition can be written as
\begin{multline}
\Bigg\{ 9 L^3
(\tKin{00})^2 \lambda_0^{(0)}
+ 9 \widetilde {\mathcal K}^2_{2,00} \frac{2 \pi a}{m^3} \frac{\mathcal K x}{(2 \pi)^6} + 9 L^3 \sum_{\vec k} \tMi{00}\slashed G_{0k}  \tMi{kk} \slashed F^{i \epsilon}_{kk}  
  \tMi{kk}  \slashed G_{k0} \tMi{00}\\
-   \frac{1}{L^3} \sum_{\vec k_1 \neq 0} \Xi_1(\vec k_1)  
-   \frac{1}{L^6} \sum_{\vec k_1, \vec k_2 \neq 0} \Xi_2(\vec k_1, \vec k_2)
\Bigg\} \Bigg\vert_{E=3m+\DEth} 
=\Mthr  +\mathcal O(1/L) \,.
\label{eq:qcondMthrL}
\end{multline}
We use Eq.~(\ref{eq:qcondMthrL}) in the following subsection to derive the
threshold expansion. 

We now return to the demonstration of Eqs.~(\ref{eq:Mthrident1})--(\ref{eq:Mthrident3}).
We first note that, in all three expressions, we can replace $E=3m+\DEth$ in the first terms with simply $E=3m$.
This is because there are no contributions to these terms that are enhanced by the tuning of $\Delta E$.
Thus shifting the energy away from threshold by $\DEth$ leads only to corrections suppressed by $1/L^3$.
The net result is that all terms in Eqs.~(\ref{eq:Mthrident1})--(\ref{eq:Mthrident3}) can be evaluated at threshold.
Note that for this it is important that the left-hand sides contain $\slashed G$ rather than $G$, since the latter
diverges at threshold.

Consider first Eq.~(\ref{eq:Mthrident1}). 
Following the arguments of Sec.~\ref{sec:KFGscaling}, the high-momentum part of the
sums on the left-hand side leads to a contribution scaling as $L^0$, in which we expect
the sums can be replaced by integrals. If any of the sums are restricted to low momenta,
then the scaling arguments of Sec.~\ref{sec:KFGscaling} can be used to show that the
contribution falls as $L\to\infty$. For example, if all the momenta are small, then,
using the result that the dominant terms in $\slashed G$ scale as $1/L$, the overall
scaling is as $L^{3-(1+n)}=L^{2-n}$, which is subleading for $n\geq3$.
Since all intermediate momenta must be large, we can restrict the sums to run over only
nonzero values without making an error when $L\to\infty$.
Doing so allows us to replace  $\slashed G$ with $\tilde G$ in the sums.
We can further replace $\tilde G$ with $G^\infty$ and $\tM$ with $\M$, as long
as we take into account all the factors of $2\omega$, $Q$ and $L^3$.
Doing so we find that 
the left- and right-hand sides of Eq.~(\ref{eq:Mthrident1}) are simply the sum and
integral, respectively, of the same summand/integrand, up to subleading corrections.\footnote{%
We also need the result that $I_n=9I_n^{(u,u)}$ at threshold.}
Thus we can rewrite (\ref{eq:Mthrident1}) as
\begin{equation}
\label{eq:Insumintdiff}
\Bigg\{
\bigg [ \frac{1}{L^3} \sum_{\vec k_1\neq 0} \cdots  \frac{1}{L^3} \sum_{\vec k_n \neq 0} 
- \int_{\vec k_1} \cdots \int_{\vec k_n} \bigg ] 9 i \M(0) i G^\infty(0, \vec k_1) \cdots i G^\infty(\vec k_n,0) \M(0) 
\Bigg\}
\Bigg\vert_{E=3m}
= \mathcal O(1/L) \,.
\end{equation}
We know from Appendix~\ref{app:Inarefin} that, although the integrand diverges in the IR,
the singularity is integrable. We also know that the integrand is nonsingular in
the high-momentum region, and is UV convergent.
Thus we can use the general result of Ref.~\cite{Luscher:1986n2} 
that such sum-integral differences vanish as a power of $1/L$.
This completes the demonstration of Eq.~(\ref{eq:Mthrident1})

Turning to Eq.~(\ref{eq:Mthrident2}), the argument proceeds along similar lines.
We can again replace $\slashed G$ with $G$ as long as we do not allow either
of the intermediate momenta to vanish. Here this is an identity, which follows because
$G_{000;0\ell m}=0$ if $\ell\ne 0$. Then we can manipulate the equation into the form
\begin{multline}
\lim_{\delta \rightarrow 0}  \bigg [ \frac{1}{L^6} \sum_{\vec k_1, \vec k_2 \neq 0}  
- \int_{\delta}   \frac{d^3 k_1}{(2 \pi)^3}    \int_{\delta}   \frac{d^3 k_2}{(2 \pi)^3} \bigg ] 
\Bigg\{ 9 i \M(0) i G^\infty(0, \vec k_1) i \M(\vec k_1) i G^{\infty}(\vec k_1, \vec k_2) 
i \M(\vec k_2) i G^\infty(\vec k_2,0) i \M(0) \\ - i \Xi_2(\vec k_1, \vec k_2) \Bigg\}\Bigg\vert_{E=3m}
= \mathcal O(1/L) \,.
\end{multline}
Here the first term in curly braces does lead to an IR divergent integral, and, correspondingly, a low-momentum
contribution to the sum that is of ${\cal O}(L^0)$,  but both of these are canceled by the $\Xi_2$ term.
Thus the expression in curly braces is integrable and nonsingular, so that the sum-integral
difference vanishes as $L\to\infty$.

The argument for Eq.~(\ref{eq:Mthrident3}) is essentially the same. Again we can replace $\slashed G$ with
$G$ as long as the intermediate sum avoids $\vec k_1=0$. The equation can then be manipulated into 
\begin{equation}
 \lim_{\delta \rightarrow 0}  
 \bigg[ \frac{1}{L^3} \sum_{\vec k_1  \neq 0} -\int_\delta \frac{d^3 k_1}{(2\pi)^3}\bigg]
   \Big\{ 9 i \M(0) i G^\infty(0, \vec k_1) i \M(\vec k_1) i G^{\infty}(\vec k_1, 0)  i \mathcal M_2(0) - i \Xi_1(\vec k_1) 
   \Big\}\Big\vert_{E=3m} =
\mathcal O(1/L) \,.
\end{equation}
Note that this equation is defined with $E=3m$ whereas the first term
in Eq.~(\ref{eq:Mthrident3}) is evaluated at 
$E=3m+\Delta E_{\text{th}}$. 
It is for this reason that the final term is needed on
the left-hand side of Eq.~(\ref{eq:Mthrident3}): it cancels an
$\mathcal O(L^0)$ contribution that arises when the first term of
(\ref{eq:Mthrident3}) is expanded about $\Delta E_{\text{th}}=0$. Only
with this term removed can we cast (\ref{eq:Mthrident3}) into the form
of a sum-integral difference as shown here. Once again, the IR
singularities cancel, by construction, in the expression in curly
braces, so the sum-integral difference vanishes as $L\to\infty$.

\subsection{Solution to the quantization condition}

\label{sec:solveDeltaE}

In this section we determine the coefficients $a_n$ in the threshold expansion of $\DEth$,
Eq.~(\ref{eq:DeltaEexp}),
by enforcing the quantization condition, Eq.~(\ref{eq:qcondMthrL}).
As noted above, we must tune $\Delta E$ to cancel the ${\cal O}(L^3)$,
${\cal O}(L^2)$ and ${\cal O}(L)$ contributions on the left-hand side of this condition.
To do so, we need the result for the $1/L$ expansion of $\lambda_0^{(0)}$, given
in Eq.~(\ref{eq:lambda0scaling}).
The algebraic manipulations needed are straightforward but tedious and we quote only the final results.

The ${\cal O}(L^3)$ and ${\cal O}(L^2)$ contributions to the left-hand side of the quantization condition
come only from the ${\cal O}(L^0)$ and ${\cal O}(1/L)$ parts of $\lambda_0^{(0)}$.
Thus these two parts must vanish.
Using Eq.~(\ref{eq:lambda0scaling}) we see that canceling 
the ${\cal O}(L^0)$ part of $\lambda_0^{(0)}$ requires
\begin{equation}
a_3 = \frac{12\pi a}{m}\,.
\label{eq:a3res}
\end{equation}
This is three times the corresponding coefficient
for two particles, which is the expected ratio
as there are now three pairs that can interact,
and is indeed the result found in Refs.~\cite{Huang,Beane2007,Tan2007,ourpt}.
We emphasize that both $\tilde F_{00}$ and $\tilde G_{00}$
contribute to $a_3$, showing the necessity of both
terms even at leading order. 

At next order,  the cancellation requires
\begin{equation}
\frac{a_4}{a_3} = - \frac{a {\cal I}}{\pi}\,.
\label{eq:a4res}
\end{equation}
This is the same {\em relative} correction as for the two-particle
case, and agrees with the results of Refs.~\cite{Huang,Beane2007,Tan2007,ourpt}.

To proceed one order higher we must determine the leading $\mathcal O(L)$ contribution from 
the sum over $\Xi_1$, a quantity whose definition is given in Eq.~(\ref{eq:Xi1def}).
We find\footnote{%
Here we are using the definition ${\cal J}= \sum_{\vec n\ne 0} 1/(\vec n^2)^2$, with
$\vec n$ a vector of integers. This is equivalent to the definition given in Appendix~\ref{app:tF00}.}
\begin{align}
\frac{1}{L^3} \sum_{\vec k_1 \neq 0} \Xi_1(\vec k_1) &
=  -  \frac{ 48^2 m^2 a^3 \mathcal J}{\pi} L  
 -9 \frac{ [32 m \pi a]^3}{(2m)^3}   m^2  
\frac{1}{L^3} \sum_{\vec k_1\neq 0} \left [   \frac{H(\vec k_1)^2-1}{k_1^{\,4}}  
+  a \frac{\sqrt{3}}{2}     \frac{H(\vec k_1)^3}{k_1^{\,3}} \right ]  
\,.
\label{eq:X1sum}
\end{align}
As we  show below, the second term on the right-hand side is of $\mathcal O(L^0)$. 
Combining the ${\cal J}$ term from $\Xi_1$ with that from Eq.~(\ref{eq:lambda0scaling}) 
we find that canceling the $\mathcal O(L)$ terms in Eq.~(\ref{eq:qcondMthrL}) requires
\begin{equation}
\frac{a_5}{a_3} =  \frac{a^2}{\pi^2}\left({\cal I}^2+{\cal J}\right)\,.
\label{eq:a5res}
\end{equation}
Again this agrees with Refs.~\cite{Huang,Beane2007,Tan2007,ourpt}.
We note that the ${\cal J}$ term in this result arises both
from $\tilde F_{00}$ and from the factors of $\tilde G_{0k}$ contained
in the sum over $\Xi_1$. Thus the agreement provides a more stringent test
of our formalism.

To determine the final coefficient, $a_6$, we must work out the $L\to\infty$ limits of all the
contributions on the right-hand side of the quantization condition (\ref{eq:qcondMthrL}).
We first consider the combination of the term containing
$\lambda^{(0)}_0$ with the ${\cal O}(L)$ contribution from $\Xi_1$.
Our tuning of $\Delta E$ has made this combination of ${\cal O}(L^0)$.
Explicitly, if we substitute the first three orders of $\DEth$ into the expression for $\lambda^{(0)}_0$
we find
\begin{multline}
\lim_{L\to\infty}\Bigg\{
9 \tKin{00}^2 L^3   \lambda_0^{(0)} + 9 \widetilde {\mathcal K}^2_{2,00} \frac{2 \pi a}{m^3} \frac{\mathcal K x}{(2 \pi)^6}+ \frac{48^2 m^2 a^3 \mathcal J}{\pi } L \Bigg\}
\Bigg\vert_{E=3m+a_3/L^3+a_4/L^4+a_5/L^5} =\\ 
\frac{576}{\pi^2} \left [ a^4 m^2 \left ( -  \mathcal{I}^3  +  \mathcal{I} \mathcal{J}  +15   \mathcal{K} \right )-  \pi^3  m^2 a \frac{a_6(L)}{a_3} +3  \pi ^4 a^2+6 \pi ^4 m^2 a^3   r  \right ] \,.
 \end{multline}
Combining this result with the remaining terms in Eq.~(\ref{eq:qcondMthrL}), which are
worked out in Appendix~\ref{app:finite},
and  demanding that the equality hold at $\mathcal O(L^0)$ then gives the expression for $a_6(L)$. 
We find
\begin{multline}
\frac{a_6(L)}{a_3} = \left(\frac{a}{\pi}\right)^3
\left[ -{\cal I}^3 + {\cal I}{\cal J} + 15 {\cal K} 
+\frac{16 \pi^3}{3} (3\sqrt 3 - 4\pi) \log\left(\frac{mL}{2\pi}\right) + \mathcal C_F+ \mathcal C_4+\mathcal C_5 \right] \\+ \frac{64\pi^2 a^2 }{m} \mathcal C_3+\frac{3\pi a}{m^2} + 6\pi ra^2
- \frac{ \mathcal M_{3,\mathrm{thr}}}{48 m^3 a_3}  \,.
\label{eq:a6res}
\end{multline}
Numerical values for the new constants $\mathcal C_F$, $\mathcal C_3$, $\mathcal C_4$ and
$\mathcal C_5$ are given in Appendix~\ref{app:finite}, while those for $\mathcal I$, 
$\mathcal J$ and $\mathcal K$ are given in Appendix~\ref{app:tF00}.
This completes our calculation of $\DEth$ through $\mathcal O(1/L^6)$. Together with results for
$a_3$, $a_4$ and $a_5$ given, respectively, in Eqs.~(\ref{eq:a3res}), (\ref{eq:a4res}) and (\ref{eq:a5res}), 
this is the main result of the paper. 

In $a_6$ only the logarithmic dependence on $L$ can be compared to that found by
the nonrelativistic calculations of Refs.~\cite{Beane2007,Tan2007}, and indeed it agrees.
The $L$-independent constants cannot be compared, 
both because relativistic effects enter at this order
and because the nonrelativistic calculations use 
different definitions for the three-particle threshold amplitude.\footnote{%
Nevertheless, we note that the $\mathcal I^3$, $\mathcal I \mathcal J$ and $\mathcal K$ terms do agree with those found by Ref.~\cite{Beane2007}.}
It is for this reason that we have carried out an independent calculation of the threshold expansion
in relativistic $\lambda \phi^4$ theory, working to cubic order~\cite{ourpt}.
Since $a$ and $a^2 r$ are both of ${\cal O}(\lambda)$ in this theory, while $\Mthr= \mathcal O(\lambda^2)$,
this allows us to check the last four terms on the right-hand side of Eq.~(\ref{eq:a6res}).
We find complete agreement. This check also shows how our definition of $\Mthr$ works in detail through
one-loop order. The remaining terms in $a_6$, i.e.~the constants on the right-hand side containing the
factor of $a^3$, have so far not been checked independently. This would require a fourth-order calculation in
the $\lambda \phi^4$ theory.

We close this section by commenting on the cutoff dependence of the various quantities in
Eq.~(\ref{eq:a6res}). The constants $\mathcal C_3$, $\mathcal C_4$ and $\mathcal C_5$ depend on
the choice of cutoff function $H$, 
as does the argument of the logarithm (though not its coefficient).
The energy of a physical finite-volume state should not depend on $H$, 
and indeed this ``scheme dependence''
is canceled by that of $\Mthr$. 
This can be seen explicitly by going back to the definition of
$\Mthr$, Eq.~(\ref{eq:Mthrdef}), in which the dependence on $H$ enters through the functions
$\Xi_1$ and $\Xi_2$, in exactly the same way as on the left-hand side of the quantization condition
(\ref{eq:qcondMthrL}). 

\section{Conclusions}
\label{sec:conc}

In this paper, we have shown how to expand the energy of the state closest to threshold 
for three interacting particles in powers of $1/L$, starting from the quantization condition derived in Refs.~\cite{us,KtoM}.
This turns out to be quite involved, but also provides insight into the workings of the formalism.
We find that the first three nontrivial terms, $a_3$, $a_4$ and $a_5$, 
as well as the logarithmic dependence in $a_6(L)$ agree with those
found previously in calculations using NRQM~\cite{Huang,Beane2007,Tan2007}. 
For a check of the volume-independent part of $a_6(L)$
(where relativistic corrections and the ambiguity in the 
definition of the three-particle threshold amplitude enter)
we have compared to a perturbative calculation in relativistic 
$\lambda \phi^4$ field theory~\cite{ourpt}.

The two-particle version of the threshold expansion~\cite{Luscher:1986n2} has been successfully used
in many numerical simulations of lattice field theories to determine the scattering 
length $a$.
Using the formula presented here, this can, in principle, be extended to the determination of 
the (suitably subtracted) three-particle scattering amplitude at threshold.
This will require accurate calculations for several volumes of
both the two- and three-particle threshold energy shifts; the former needed to determine $a$ and the
effective range $r$. 
This will be challenging in practice,
since one must control the volume dependence up to ${\cal O}(1/L^6)$.

The development of the threshold expansion for three particles is much more challenging than 
in the two-particle case.\footnote{%
The latter is given up to ${\cal O}(1/L^5)$ in Ref.~\cite{Luscher:1986n2} and
to one higher order in Appendix C of Ref.~\cite{ourpt}.}
The main reason for this difficultly is that the matrices appearing in the quantization condition cannot be truncated
when one works at ${\cal O}(1/L^6)$. While this added to the technical challenge, it also led to the conversion
of the unphysical quantity $\Kdf$, which appears in the quantization condition, into the physical subtracted
threshold amplitude $\Mthr$. This was essential for the final result for $\DEth$ to depend only on physical
quantities.

One might be concerned that the complications that we have had to deal with here will carry over 
to the practical application of the three-particle quantization condition.
This is not, however, the case. When one does a $1/L$ expansion one loses one of the
key simplifying features of the quantization condition. 
This feature, stressed in Ref.~\cite{KtoM}, is that, if one truncates the two-particular angular momentum space, 
then the matrices of the determinant condition, Eq.~(\ref{eq:qcond}), become finite. 
This is because the remaining matrix index, $\vec k = 2 \pi \vec n_k/L$, 
is automatically truncated by the smooth cutoff function $H(\vec k)$. 
In particular $H(\vec k)$ vanishes for $k \gtrsim m$ implying
that $n_k$ is constrained to satisfy $2 \pi n_k \lesssim m L$. 
Thus only a finite number of values of $\vec n_k$ need be used when applying this formalism to
numerical simulations.
By contrast, the threshold expansion must be valid for arbitrarily large $L$, which implies
that all $\vec n$ can contribute. 

\section*{Acknowledgments}
The work of SS was supported in part by the United States Department of Energy 
grant DE-SC0011637.

 \appendix

\section{Evaluation of $\tilde F_{00}$}
\label{app:tF00}

In this appendix we expand the quantity
$\tilde F_{00}$ in powers of $1/L$ taking $\Delta E=E-3m$ to
scale as $1/L^3$.
We recall that $\tilde F_{00}\equiv \tilde F_{000;000}$,
with the latter defined in Eq.~(\ref{eq:Ftilexplicit}).
For the analysis in the main text,
we need to keep terms in $\tilde F_{00}$ up to ${\cal O}(1/L^3)$.

We start from the expressions given in Eqs.~(\ref{eq:Ftozeta}) and (\ref{eq:zetadef}).
As the spectator momentum is $\vec k=0$, 
the scattered pair are already in their CM frame, so the boost factor
$\gamma$ in Eq.~(\ref{eq:E20q0}) is unity. It follows that $\vec r=\vec n_a$.
Thus we obtain
\begin{align}
\tilde F_{00} &= \frac{1}{16m \omega_{q}} \left\{  
\frac{1}{q^2 L^3} +
\frac{1}{4\pi^2 L}
\left[\sum_{\vec n_a \neq 0} -\widetilde{\rm PV} \int_{\vec n_a}\right]
\frac{H(\vec a)^2}{x^2-n_a^2} \right\}
\,,
\label{eq:F00red}
\end{align}
where $x=q L/(2\pi)$ and we have used the fact that  $\vec b_{ka}=-\vec a$,
and the evenness of $H(\vec a)$, to rewrite the regulator function.
We have also absorbed the $\rho$ term in Eq.~(\ref{eq:Ftozeta})
into the integral over $\vec n_a$ by reverting to the $\widetilde{\rm PV}$ pole prescription.
As explained in Ref.~\cite{us}, this prescription leads to
integrals such as that in Eq.~(\ref{eq:F00red}) being real and smooth
functions of $x^2$. In particular, the cusp at
$x^2=0$ present with the $i\epsilon$ prescription is absent
with the $\widetilde{\rm PV}$ prescription.

In Eq.~(\ref{eq:F00red}), we have pulled out the $\vec n_a=0$ term from the sum 
since this scales as $L^0$ [using Eq.~(\ref{eq:E20q0})].
The remainder scales as $1/L$, as already discussed in Sec.~\ref{sec:KFGscaling}.
For the sum in Eq.~(\ref{eq:F00red}) we can use the fact that
$|x^2|\sim 1/L \ll n_a$ (and the absence of the $\vec n_a=0$
term in the sum) to expand the summand in powers of $x^2$, leading to
\begin{equation}
\sum_{\vec n_a\ne 0} \frac{H(\vec a)^2}{x^2-n_a^2}
= - 
\sum_{j=0}^\infty
\left[\frac{q^2 L^3}{4\pi^2 L}\right]^j 
\sum_{\vec n_a\ne 0} \frac{H(\vec a)^2}{(n_a^2)^{1+j}} 
\,.
\label{eq:F00sumexp}
\end{equation}
Although $H$ is needed to regulate the UV only for $j=0$,
we cannot drop it from the other terms, as doing so leads to
potential power law corrections. 
To see this, we note that [using Eq.~(\ref{eq:qkstardef})]
\begin{equation}
\frac{E_{2,a}^{*2}}{4m^2} 
= \frac52 - \frac32 \sqrt{1+ \frac{a^2}{m^2}} 
+ {\cal O}\left(\frac{\Delta E}{m}\right)\,,
\end{equation}
implying that the regulator function takes the explicit form
\begin{equation}
H(\vec a)
= J\left(\frac52 - \frac32 \sqrt{1+ \frac{n_a^2}{N_{\rm cut}^2}}\right) 
+ {\cal O}\left(\frac{\Delta E}{m}\right)
\label{eq:Hares}
\end{equation}
with
\begin{equation}
N_{\rm cut} = \frac{mL}{2\pi}\,.
\end{equation}
Given the definition of the function $J$, Eq.~(\ref{eq:HJdef}),
this implies that the sum over $\vec n_a$ is cut off (smoothly) at
$n_a \approx N_{\rm cut}$. Since this cutoff depends on $L$, it
can introduce further $L$ dependence  
in the individual terms of Eq.~(\ref{eq:F00sumexp}). For example, in the
sum over $H(\vec a)^2/n_a^4$, it is easy to see that the cutoff
leads to a $1/(mL)$ correction. Since this sum arises in a $1/L^2$ term
in $\tilde F_{00}$, the correction would enter at ${\cal O}(1/L^3)$,
which is the highest order that we are controlling. Thus we cannot remove
the cutoff at this stage.\footnote{%
We can, however, drop the $\Delta E$ term in Eq.~(\ref{eq:Hares}), since
this is proportional to $1/L^3$, pushing the total
power to $1/L^5$, i.e. beyond the order we are working.}

We would like to do a similar expansion in powers of $x^2$ for the
integral in Eq.~(\ref{eq:F00red}). We know that this must be possible since the
$\widetilde{\rm PV}$ prescription leads to smooth, nonsingular dependence
on $x^2$, including at $x^2=0$. Naively expanding, however, leads to
integrals that diverge at $\vec n_a=0$. To proceed, we first pull out the $x^2=0$ term
\begin{align}
\widetilde{\rm PV} \int_{\vec n_a} \frac{H(\vec a)^2}{x^2-n_a^2}
&= -\int_{\vec n_a} \frac{H(\vec a)^2}{n_a^2}
+ \widetilde{\rm PV} \int_{\vec n_a} 
\frac{x^2 H(\vec a)^2 }{n_a^2(x^2-n_a^2)}\,,
\label{eq:PViden1}
\end{align}
where no pole prescription is needed for the IR and UV convergent
integral in the first term on the right-hand side.
Next we use the result
\begin{equation}
\widetilde{\rm PV}\int_{\vec n_a}
\frac{1 }{n_a^2(x^2-n_a^2)} = 0\,,
\label{eq:PVtildeident}
\end{equation}
which can be shown by explicit computation.
Note that this integral is finite both in the UV and IR
for $x^2\ne 0$, so that no regulation is required, and the $x^2=0$
result is obtained by smoothness.
Subtracting this vanishing integral from that appearing in the second
term on the right-hand side of Eq.~(\ref{eq:PViden1}) we find 
\begin{align}
\widetilde{\rm PV} \int_{\vec n_a} \frac{x^2H(\vec a)^2 }{n_a^2(x^2-n_a^2)}
&=
x^2\; \widetilde{\rm PV}\! \int_{\vec n_a} \frac{H(\vec a)^2 -1}{n_a^2(x^2-n_a^2)} \,,
\label{eq:PViden2}
\\
&=
- x^2 \sum_{j=0}^\infty (x^2)^j
\int_{\vec n_a} \frac{H(\vec a)^2 -1}{(n_a^2)^{2+j}}
\label{eq:PViden3}
\,.
\end{align}
Here we are allowed to do a Taylor expansion because the resulting
integrals are convergent both in the IR and UV. The IR convergence is
assured by the factor of $H(\vec a)^2-1$, a function of $n_a^2$ 
all of whose derivatives vanish at $n_a=0$. The UV convergence is
manifest for all $j$. Once again, despite the UV convergence, we cannot
drop the factors of $H$ since they give rise to power law corrections.
Finally we note that no pole prescription is needed in the integrals
in Eq.~(\ref{eq:PViden3}).

Collecting these results we obtain the $1/L$ expansion for $\tilde F_{00}$:
\begin{equation}
\tilde F_{00}  =   \frac{1}{16 m \omega_q} 
\frac{1}{q^2 L^3} \left\{ 1 - 
\sum_{j = 1}^{\infty} \left [ \frac{q^2 L^3}{4 \pi^2 L} \right ]^j 
\mathcal I_{j}   \right \} \,.
\label{eq:F00res}
\end{equation}
Here
\begin{equation}
\mathcal I_{j} = \begin{cases}
\left[\sum_{\vec n_a \neq 0} -\int_{\vec n_a}\right]
\frac{H(\vec a)^2}{\vec n_a^2} 
& j=1\,, \\
\sum_{\vec n_a\ne 0} \frac{H(\vec a)^2}{(\vec n_a)^{2j}} - 
\int_{\vec n_a} \frac{H(\vec a)^2-1}{(\vec n_a)^{2j}}
& j\ge 2\,.
\label{eq:Imres}
\end{cases}
\end{equation}
These quantities retain an implicit dependence on $L$ through the
cutoff functions. However, this dependence is expected to be
exponentially suppressed (falling as $\exp(-N_{\rm cut})$), since
in the derivation of the formalism in Ref.~\cite{us} the dependence
on the form of $H$ is exponentially suppressed.
Indeed, it is simple to check that the leading power law dependence on
$N_{\rm cut}$ cancels between the sums and integrals for ${\cal I}_m$
with $m\ge 2$. Furthermore, numerically evaluating the expressions, 
we observe that the convergence as $N_{\rm cut}$ increases is rapid and
consistent with exponential. Thus we can replace these quantities
with their values when $N_{\rm cut}\to \infty$.
In the notation of Ref.~\cite{Beane2007} the first three become
\begin{equation}
\mathcal I_1 \xrightarrow[N_{\rm cut}\to \infty]{} \mathcal I\,,\quad
\mathcal I_2 \xrightarrow[N_{\rm cut}\to \infty]{} \mathcal J\,,\quad
\mathcal I_3 \xrightarrow[N_{\rm cut}\to \infty]{} \mathcal K\,.
\label{eq:IJK}
\end{equation}
We have checked that the numerical values we obtain for
$\mathcal I$, $\mathcal J$ and $\mathcal K$ agree with those
quoted (to about 12 significant figures) in Ref.~\cite{Beane2007}.\footnote{%
Indeed, for $\mathcal I_2$ and $\mathcal I_3$, the expressions in
Eq.~(\ref{eq:Imres}) provide a numerically 
efficient way of evaluating the sums.}
Quoting only four decimal places, the values are
$\mathcal I=-8.914$, $\mathcal J=16.532$ and $\mathcal K=8.402$.
Making the replacements of Eq.~(\ref{eq:IJK}) we obtain the result
(\ref{eq:tildeF00final}) quoted in the main text.

\section{Proof that $I_{n>3}$ are finite at threshold}
\label{app:Inarefin}

In this appendix we prove that, for $n \geq 3$, the integrals $I_n(\vec p, \hat a'^*; \vec k, \hat a^*)$, 
defined in Eqs.~(\ref{eq:Indef}) and (\ref{eq:symmdef}), are finite at threshold, $E =3m$. 
The potential divergence is only in the infrared, since the functions $H$ contained
in $G^\infty$ [defined in Eq.~(\ref{eq:Ginfdef})] regulate the ultraviolet.
As will become clear in the following, the divergences in any $I_n$ occur only when the
external spectator momenta are set to $\vec p=\vec k=0$,
so we primarily consider this case. Setting $\vec p=\vec k=0$ at threshold implies 
in turn that $\vec a'^*= \vec a^*=0$, so that the $I_n$ are pure $s$-wave, with
no dependence on $\hat a'^*$ and $\hat a^*$.

When all momenta (both external and internal) are in the IR regime, $k\ll m$,
the energy denominators  in each factor of $G^\infty$ take their nonrelativistic form
\begin{equation}
E- \omega_k-\omega_p-\omega_{pk} + i \epsilon
\longrightarrow
- [\vec k^2 + \vec p^2 + (\vec k+\vec p)^2 - i \epsilon]/(2m)
\label{eq:B1}
\end{equation}
Thus, if we set the external momenta to zero, and collect the $n$ three-vectors
that are being integrated into a $3n$-dimensional vector
$\vec Q \equiv (\vec k_1, \cdots, \vec k_n)$, 
we have (since $I_n$ contains $n+1$ factors of $G^\infty$ and $n$ integrals)\footnote{%
Each factor of $G^\infty$ contains a double sum over angular momentum indices,
but we consider here only $s$-wave contributions, since these dominate in the IR 
due to the factor of  $(k^*)^{\ell'} (p^*)^\ell\sim Q^{\ell'+\ell}$ in $G^\infty$.}
\begin{equation}
I_n \sim \int \! d Q \int \! d  \Omega \  Q^{3n-1} \frac{1}{Q^{2(n+1)} f(\Omega)} 
\propto \int d Q\; Q^{n-3} \,.
\label{eq:B2}
\end{equation}
Here $\Omega$ stands for the collective angular coordinates. 
Thus the integral is IR divergent by power-counting for $n=1$ and $2$,  
while finite for $n\geq 3$. 
There is, however, another possible source of divergence, namely that
$f(\Omega)$ can have zeroes. 
These occur when some, but not all, of the $G^\infty$ factors diverge. 
It turns out, however, that these zeroes result in no additional divergences since 
they are canceled by corresponding zeroes in the numerator. 
Thus the naive overall power-counting result is correct.

To explain this, we first replace $I_n$ (with vanishing external momenta)
with the simpler integral
\begin{equation}
I_{n,\mathrm{IR}} \equiv \int_{\vec k_1, \cdots, \vec k_n} 
\frac{1}{2\vec k_1^2}
 \frac{1}{\vec k_1^2 + \vec k_2^2 + (\vec k_1 + \vec k_2)^2} 
\cdots 
\frac{1}{\vec k_{n-1}^2 + \vec k_n^2 + (\vec k_{n-1} + \vec k_n)^2} 
\frac{1}{2 \vec k_n^2} \,.
\end{equation}
This removes extraneous factors while maintaining the IR properties of the integral.
We have dropped factors of $i\epsilon$ since they are not needed to regularize
these integrals when working at threshold.

Next we consider the $n=3$ case in detail.
\begin{align}
I_{3,\mathrm{IR}} & \equiv \int_{\vec k_1, \vec k_2, \vec k_3} 
\frac{1}{2\vec k_1^2}
 \frac{1}{\vec k_1^2 + \vec k_2^2 + (\vec k_1 + \vec k_2)^2} 
  \frac{1}{\vec k_{2}^2 + \vec k_3^2 + (\vec k_{2} + \vec k_3)^2} 
  \frac{1}{\vec 2 k_3^2} 
  \label{eq:B4}
  \,, \\
& = \frac{1}{8 (2 \pi)^6} \int d k_1 \int d k_2 k_2^2 \int d \cos \theta_{12} 
\int d k_3 \int d \cos \theta_{23} 
 \frac{1}{  k_1^2 +  k_2^2 +   k_1  k_2 \cos \theta_{12}} 
  \frac{1}{ k_{2}^2 +  k_3^2 +  k_{2}  k_3 \cos \theta_{23}}
   \,, \\
& = \frac{1}{8 (2 \pi)^6}  \int d Q \int_0^{\pi/2} d \phi   
\int_0^{\pi/2} d \theta   \int_0^\pi d \theta_{12} \int_0^\pi d \theta_{23}   
\nonumber \\
& \times  \frac{\sin \theta_{12} \sin \theta_{23} \sin^3 \theta \sin^2 \phi}
{ \left (   \sin^2 \theta +     \sin^2 \theta \sin \phi \cos \phi  \cos \theta_{12}   \right )  
\left ( \sin^2 \theta \sin^2 \phi + \cos^2 \theta + \sin \theta \sin \phi \cos \theta \cos \theta_{23} \right )} 
\,.
\end{align}
Here we are using the variables
$(k_1,k_2,k_3)= Q(\sin\theta\cos\phi,\sin\theta\sin\phi,\cos\theta)$.
The lack of divergence in the overall $Q$ integral agrees with our analysis above.
One of the possible divergences in the angular integrals occurs when $\theta\approx 0$
(corresponding to $\vec k_1$ and $\vec k_2$ vanishing but not $\vec k_3$).
In this limit, the integrand becomes
\begin{equation}
 \frac{\sin \theta_{12} \sin \theta_{23}   \sin^2 \phi \ \ \theta^3}{ \left (  1 +      \sin \phi \cos \phi  \cos \theta_{12}   \right ) \ \ \theta^2   }  + \mathcal O(\theta^2) \,,
\end{equation}
so the integral over $\theta$ is finite.
There is a similar possible divergence
when $\phi\approx 0$, $\theta\approx \pi/2$
(corresponding to $\vec k_2$ and $\vec k_3$ vanishing but not $\vec k_1$),
but it is clear from the symmetry of the original expression (\ref{eq:B4}) under
$\vec k_1 \leftrightarrow \vec k_3$ that this will also lead to a convergent integral.
Finally, the divergences when $\vec k_1$ and/or $\vec k_3$ both vanish
(but not $\vec k_2$) are manifestly integrable.

An alternative way of stating this result  is that, when any pair of momenta vanish,
there are two measure factors of $k^2$ and two denominators vanishing as $k^2$,
so the IR divergence cancels.
In this form, the argument is easily generalized to all $I_n$ with $n \geq 3$.
If $j$ coordinates vanish there will be $j$ measure factors of $k^2$ and,
at most, $j$ denominators vanishing as $k^2$. 
(To achieve this number of diverging denominators the momenta must be sequential 
and include either the first or last momenta.)
Thus all subintegrals are IR convergent, and we deduce that $I_n$ itself is finite.

The discussion so far assumes that both external momenta are set to zero. 
If one (or both) are nonvanishing, then it is straightforward to see that the
loss of one (or two) potentially vanishing denominators is sufficient to make 
$I_n$ IR finite for all $n>0$, including $n=1$ and $2$. 
This assumes that $E$ is evaluated at threshold.
Similarly, all $I_n$ are IR finite if any of the internal angular momenta are taken
to be anything other than $s$-wave. For example, in $I_1$, whose overall IR
divergence is linear [$\int dQ Q^{-2}$ from Eq.~(\ref{eq:B2})],
choosing the internal $\M$ to be in a $p$-wave leads to an extra $Q^2$
(one factor of $Q$ from each of the adjacent $G^\infty$)
and removes the divergence.

\section{Calculation of finite terms}
\label{app:finite}

In this appendix we calculate the contributions of $\mathcal O(L^0)$ arising from the second, third and fourth
terms on the left-hand side of the quantization condition Eq.~(\ref{eq:qcondMthrL}).
These are  needed in Sec.~\ref{sec:solveDeltaE} to find the coefficients in the expansion of
the threshold energy $\DEth$.

We begin with
\begin{align}
{\cal X}_F &=
\lim_{L\to\infty}
\Bigg\{ 9 L^3 \sum_{\vec k} \tMi{00}\slashed G_{0k}  \tMi{kk} 
\slashed F^{i \epsilon}_{kk}    \tMi{kk}  \slashed G_{k0} \tMi{00} \Bigg\}\Bigg\vert_{E=3m+\DEth}
\\
&=
\lim_{L\to\infty}
\Bigg\{ 9 L^3 \sum_{\vec k \ne 0} \tMi{00}\tilde G_{0k}  \tMi{kk} 
\tilde F^{i \epsilon}_{kk}    \tMi{kk}  \tilde G_{k0} \tMi{00} \Bigg\}\Bigg\vert_{E=3m}
\,.
\end{align}
We recall that the notation here indicates that only $s$-wave contributions are kept.
In the second form we have made two changes. The first is an identity: we can replace
the slashed $G$ and $F^{i\epsilon}$ with the tilded versions as long as we remove
$\vec k=0$ from the sum. The second is to work directly at threshold, which is allowed
since the absence of the $\vec k=0$ term means that $\DEth\sim 1/L^3$ always leads
to a correction suppressed by $1/L$.

We recall from Sec.~\ref{sec:KFGscaling} that the sum is dominated by small momenta,
so it is legitimate to use nonrelativistic expansions of the various quantities worked out in
that section and keep only the leading terms. Thus $\tMi{00}$ and $\tMi{kk}$ can be replaced
by the constant $-64\pi m^2 a$ [using Eq.~(\ref{eq:K00res}) and the equality of $\tMi{00}$ and
$\tKin{00}$ at threshold]. Using Eq.~(\ref{eq:Gtildef}), the leading term in $\tilde G_{0k}$ is
given by
\begin{equation}
\tilde G_{0k}\big\vert_{E=3m} = - \frac1L \frac1{32 \pi^2 m^2} \frac{1}{n_k^2} + {\cal O}(1/L^2)\,,
\end{equation}
where $\vec k=2\pi \vec n_k/L$. Note that in the small-momentum regime we can set $H(\vec k)$ to
unity.
Finally, using Eq.~(\ref{eq:Ftozeta}), and recalling that $\tilde F^{i\epsilon}$ differs from
$\tilde F$ by dropping the $\rho$ term, we have
\begin{equation}
\tilde F_{kk}^{i \epsilon}\bigg\vert_{E=3m}
=
\frac1L \frac{1}{64\pi^2 m^2} 
\left[\sum_{\vec n_a}-\int_{\vec n_a}\right]
\frac{H(\vec a) H(\vec b_{ka})}{x^2-r^2} 
+ {\cal O}(1/L^2)\,,
\end{equation}
where $x^2=-3 n_k^2/4$,  and $\vec r$ is defined in Eq.~(\ref{eq:rdef}), except that we can set $\gamma=1$
in our kinematic regime.
Since $r^2>0$ while $x^2 <0$ there is no singularity in the summand/integrand, and thus
the $i\epsilon$ regularization can be dropped.

The sum-integral difference can be evaluated using the Poisson summation formula
\begin{align}
\left[\sum_{\vec n_a}-\int_{\vec n_a}\right]
\frac{H(\vec a) H(\vec b_{ka})}{x^2-r^2} 
&=
- \sum_{\vec s\ne 0} e^{i\pi \vec s\cdot \vec n_k}
\int d^3 r \; e^{2\pi i \vec s\cdot \vec r} 
\frac{H(\vec a) H(\vec b_{ka})}{|x|^2+r^2} 
\\
&= -\pi 
\sum_{\vec s\ne 0} e^{i\pi \vec s\cdot \vec n_k}
\frac{e^{-2\pi|x| s}}{s} + {\cal O}(e^{-mL})
\,,
\label{eq:exponentialsuppression}
\end{align}
where $\vec s$ is a vector of integers.
To obtain the second line we have used the fact that the Fourier transform in the first line
is dominated by values of $r$ satisfying $r\lesssim |x| = {\cal O}(1)$, which in turn implies
that $\vec a$ and $\vec b_{ka}$ are small, so that the cutoff functions $H$ can be replaced
by unity up to exponentially small corrections. Doing so we can evaluate the integral and
obtain the result on the second line. The result shows that the zeta-function 
(sum-integral difference) falls exponentially with increasing $|x|$.
When evaluating this expression numerically, we find that the sum converges rapidly 
for $|x| \gtrsim 1$.

Combining these results, we find that
\begin{align}
{\cal X}_F
&= \frac{576 m^2 a^4}{\pi^2}  (-4\pi)
\sum_{\vec n_k\ne 0} \frac1{n_k^4} \sum_{\vec s\ne 0} e^{i\pi \vec s \cdot \vec n_k} \frac{e^{-2\pi |x| s}}{s}
\\
&\equiv
 \frac{576 m^2 a^4}{\pi^2}\;  {\cal C}_F
\,,
\end{align}
where numerical evaluation leads to ${\cal C}_F= -0.493036$. This accuracy is obtained by summing up to
$n_k^2 =11$ and $s^2 = 12$.

\bigskip
We next evaluate the contributions coming from the sum over $\Xi_1$,
i.e.~those from the last term in Eq.~(\ref{eq:X1sum}).
These are
\begin{align}
{\cal X}_{1A} &=  \lim_{L\to\infty} \left\{ 9 \frac{ [32 m \pi a]^3}{(2m)^3}   m^2  
\frac{1}{L^3} \sum_{\vec k\neq 0}  \frac{H(\vec k)^2-1}{k^{\,4}}  \right\} \,,
\label{eq:X1Adef} 
\\
{\cal X}_{1B} &=9 \frac{ [32 m \pi a]^3}{(2m)^3}   m^2  
\frac{1}{L^3} \sum_{\vec k\neq 0}
 a \frac{\sqrt{3}}{2}     \frac{H(\vec k)^3}{k^{\,3}} \,,
 \label{eq:X1Bdef}
\end{align}
where we are implicitly working at $E=3m$ in the cutoff functions $H$. 
In the second quantity we cannot send $L \to \infty$ but we implicitly 
discard all terms which vanish as $L \to \infty$.
Recalling that the Taylor expansion of $H$ about $\vec k=0$ is unity to all orders,
we see that the summand of ${\cal X}_{1A}$ is non-singular, so that the sum can be replaced by an integral
in the $L\to\infty$ limit. This leads to the result
\begin{align}
{\cal X}_{1A} &=   576 \pi m a^3 \; 64\pi^2 \; {\cal C}_3\,,
\\
{\cal C}_3 &\equiv \int \frac{d^3 k}{(2\pi)^3} \frac{m [H(\vec k)^2\!-\!1]}{k^4} = -0.05806\,.
\end{align}
For $\mathcal X_{1B}$, the summand has a pole so the sum cannot be replaced by an integral.
Furthermore, the sum has a logarithmic divergence in the UV that is cut off by $H$ 
and leads
to a $\log(mL)$ dependence. To determine its form we rewrite the expression as
\begin{equation}
\mathcal X_{1B} = \frac{576 m^2 a^4}{\pi^2} 4 \pi^2 \sqrt3 \sum_{\vec n_k}  \frac{H(2\pi \vec n_k/L)^3}{n_k^3} 
\,.
\end{equation}
From the definition of $H$, Eq.~(\ref{eq:HJdef}), we know it vanishes when $(E_{2,k}^*)^2$ drops to zero.
From the definition of $(E_{2,k}^*)^2$ in Eq.~(\ref{eq:qkstardef}), we find (when $E=3m$) that it
vanishes when $k/m=4/3$. Thus, in terms, of $\vec n_k=(L/2\pi) \vec k$, 
the sum is cut off at $(4/3)N_{\rm cut}$ where
$N_{\rm cut}=mL/(2\pi)$.  Approximating the UV part of the sum with an integral gives the logarithmic
dependence, and by numerical evaluation we can determine the constant underneath:
\begin{equation}
\sum_{\vec n_k} \frac{H(2\pi \vec n_k/L)^3}{n_k^3} = 4\pi \log N_{\rm cut} + 1.54861 + {\cal O}(1/L)
\,.
\end{equation}
Combining these results we find
\begin{align}
\mathcal X_{1B} &= \frac{576 m^2 a^4}{\pi^2}
\left(  16 \pi^3 \sqrt3 \log N_{\rm cut} + {\cal C}_4\right) + {\cal O}(1/L)
\,,
\\
{\cal C}_4 &= 105.892\,.
\end{align}

The final contribution is that from $\Xi_2$, which is
\begin{equation}
{\cal X}_2 = \frac1{L^6} \sum_{\vec k_1,\vec k_2\ne 0} \Xi_2(\vec k_1,\vec k_2) \,.
\end{equation}
This can be evaluated at $E=3m$ (which only affects the cutoff functions 
$H$ contained in $\Xi_2$).
Using the definition of $\Xi_2$, Eq.~(\ref{eq:Xi2def}),  we find
\begin{equation}
{\cal X}_2 = \frac{576 m^2 a^4}{\pi^2} 16 \sum_{\vec n_1,\vec n_2\ne0} 
\frac{H(2\pi \vec n_1/L)^2 H(2\pi \vec n_2/L)^2}{n_1^2[n_1^2+n_2^2+(\vec n_1+\vec n_2)^2]n_2^2}
\,.
\end{equation}
Again the sum has a logarithmic UV divergence, and, pulling this out, we find by numerical evaluation that
\begin{align}
{\cal X}_2 &= \frac{576 m^2 a^4}{\pi^2} \left( \frac{64\pi^4}3 \log N_{\rm cut} - {\cal C}_5\right) + {\cal O}(1/L)
\,,\\
{\cal C}_5 &
=  1947
\,.
\end{align}
We note that, while the coefficient $\mathcal C_5$ appears large, it is approximately the same size
as the coefficient of the logarithm: $64 \pi^4/3 \approx 2080$.

\bibliography{ref} 

\end{document}